\documentclass[submission,copyright,creativecommons]{eptcs}
\providecommand{\event}{MSFP 2012} 

\makeatletter
\@ifundefined{lhs2tex.lhs2tex.sty.read}%
  {\@namedef{lhs2tex.lhs2tex.sty.read}{}%
   \newcommand\SkipToFmtEnd{}%
   \newcommand\EndFmtInput{}%
   \long\def\SkipToFmtEnd#1\EndFmtInput{}%
  }\SkipToFmtEnd

\newcommand\ReadOnlyOnce[1]{\@ifundefined{#1}{\@namedef{#1}{}}\SkipToFmtEnd}
\usepackage{amstext}
\usepackage{amssymb}
\usepackage{stmaryrd}
\DeclareFontFamily{OT1}{cmtex}{}
\DeclareFontShape{OT1}{cmtex}{m}{n}
  {<5><6><7><8>cmtex8
   <9>cmtex9
   <10><10.95><12><14.4><17.28><20.74><24.88>cmtex10}{}
\DeclareFontShape{OT1}{cmtex}{m}{it}
  {<-> ssub * cmtt/m/it}{}

\DeclareFontShape{OT1}{cmtt}{bx}{n}
  {<5><6><7><8>cmtt8
   <9>cmbtt9
   <10><10.95><12><14.4><17.28><20.74><24.88>cmbtt10}{}
\DeclareFontShape{OT1}{cmtex}{bx}{n}
  {<-> ssub * cmtt/bx/n}{}

\newcommand{\Conid}[1]{\mathit{#1}}
\newcommand{\Varid}[1]{\mathit{#1}}
\newcommand{\anonymous}{\kern0.06em \vbox{\hrule\@width.5em}}
\newcommand{\plus}{\mathbin{+\!\!\!+}}
\newcommand{\bind}{\mathbin{>\!\!\!>\mkern-6.7mu=}}
\newcommand{\sequ}{\mathbin{>\!\!\!>}}

\renewcommand{\geq}{\geqslant}
\usepackage{polytable}

\@ifundefined{mathindent}%
  {\newdimen\mathindent\mathindent\leftmargini}%
  {}%

\def\resethooks{%
  \global\let\SaveRestoreHook\empty
  \global\let\ColumnHook\empty}
\newcommand*{\savecolumns}[1][default]%
  {\g@addto@macro\SaveRestoreHook{\savecolumns[#1]}}
\newcommand*{\restorecolumns}[1][default]%
  {\g@addto@macro\SaveRestoreHook{\restorecolumns[#1]}}
\newcommand*{\aligncolumn}[2]%
  {\g@addto@macro\ColumnHook{\column{#1}{#2}}}

\resethooks

\newcommand{\onelinecommentchars}{\quad-{}- }
\newcommand{\commentbeginchars}{\enskip\{-}
\newcommand{\commentendchars}{-\}\enskip}

\newcommand{\visiblecomments}{%
  \let\onelinecomment=\onelinecommentchars
  \let\commentbegin=\commentbeginchars
  \let\commentend=\commentendchars}

\newcommand{\invisiblecomments}{%
  \let\onelinecomment=\empty
  \let\commentbegin=\empty
  \let\commentend=\empty}

\visiblecomments

\newlength{\blanklineskip}
\newcommand{\hsindent}[1]{\quad}
\let\hspre\empty
\let\hspost\empty

\EndFmtInput
\makeatother
\ReadOnlyOnce{polycode.fmt}%
\makeatletter

\newcommand{\hsnewpar}[1]%
  {{\parskip=0pt\parindent=0pt\par\vskip #1\noindent}}

\newcommand{\hscodestyle}{}

\newcommand{\sethscode}[1]%
  {\expandafter\let\expandafter\hscode\csname #1\endcsname
   \expandafter\let\expandafter\endhscode\csname end#1\endcsname}

  {\par\noindent
   \advance\leftskip\mathindent
   \hscodestyle
   \let\\=\@normalcr
   \(\pboxed}%
  {\endpboxed\)%
   \par\noindent
   \ignorespacesafterend}

  {\hsnewpar\abovedisplayskip
   \advance\leftskip\mathindent
   \hscodestyle
   \let\hspre\(\let\hspost\)%
   \pboxed}%
  {\endpboxed%
   \hsnewpar\belowdisplayskip
   \ignorespacesafterend}

  {\hsnewpar\abovedisplayskip
   \advance\leftskip\mathindent
   \hscodestyle
   \let\\=\@normalcr
   \(\pboxed}%
  {\endpboxed\)%
   \hsnewpar\belowdisplayskip
   \ignorespacesafterend}

\newcommand{\plainhs}{\sethscode{plainhscode}}

\plainhs

  {\hsnewpar\abovedisplayskip
   \advance\leftskip\mathindent
   \hscodestyle
   \let\\=\@normalcr
   \(\parray}%
  {\endparray\)%
   \hsnewpar\belowdisplayskip
   \ignorespacesafterend}

  {\parray}{\endparray}

  {\(\parray}{\endparray\)}

\def\codeframewidth{\arrayrulewidth}
\RequirePackage{calc}

  {\parskip=\abovedisplayskip\par\noindent
   \hscodestyle
   \arrayrulewidth=\codeframewidth
   \tabular{@{}|p{\linewidth-2\arraycolsep-2\arrayrulewidth-2pt}|@{}}%
   \hline\framedhslinecorrect\\{-1.5ex}%
   \let\endoflinesave=\\
   \let\\=\@normalcr
   \(\pboxed}%
  {\endpboxed\)%
   \framedhslinecorrect\endoflinesave{.5ex}\hline
   \endtabular
   \parskip=\belowdisplayskip\par\noindent
   \ignorespacesafterend}

\newcommand{\framedhslinecorrect}[2]%
  {#1[#2]}

  {\(\def\column##1##2{}%
   \let\>\undefined\let\<\undefined\let\\\undefined
   \newcommand\>[1][]{}\newcommand\<[1][]{}\newcommand\\[1][]{}%
   \def\fromto##1##2##3{##3}%
   }{\) }%

  {\let\orighscode=\hscode
   \let\origendhscode=\endhscode
   \def\endhscode{\def\hscode{\endgroup\def\@currenvir{hscode}\\}\begingroup}
   \orighscode\def\hscode{\endgroup\def\@currenvir{hscode}}}%
  {\origendhscode
   \global\let\hscode=\orighscode
   \global\let\endhscode=\origendhscode}%

\makeatother
\EndFmtInput
\usepackage{ifsym}
\usepackage{tikz}
\usetikzlibrary{matrix,arrows,fit}
\usepackage{amsmath}
\usepackage{amsthm}
\usepackage{rotating}

\usepackage{amsthm}

\theoremstyle{definition}

\title{Tracing monadic computations and representing effects}

\author{
Maciej Pir{\'o}g and Jeremy Gibbons
\institute{Department of Computer Science \\ University of Oxford}
\email{Maciej.Pirog@cs.ox.ac.uk \qquad Jeremy.Gibbons@cs.ox.ac.uk}
}
\def\titlerunning{Tracing monadic computations and representing effects}
\def\authorrunning{Maciej Pir{\'o}g \& Jeremy Gibbons}

\begin{document}

\maketitle

\begin{abstract}
In functional programming, monads are supposed to encapsulate computations, effectfully producing the final result, but keeping to themselves the means of acquiring it. For various reasons, we sometimes want to reveal the internals of a computation. To make that possible, in this paper we introduce monad transformers that add the ability to automatically accumulate observations about the course of execution as an effect. We discover that if we treat the resulting trace as the actual result of the computation, we can find new functionality in existing monads, notably when working with non-terminating computations.
\end{abstract}

\section{Introduction}

Consider a simple database in which different users store data under keys. This can be represented in Haskell as a data structure of type \ensuremath{[\mskip1.5mu (\Conid{User},[\mskip1.5mu (\Conid{Key},\Conid{Data})\mskip1.5mu])\mskip1.5mu]}. We can define a function \ensuremath{\Varid{getData}} which looks up a value for a specified user and a key.

\begin{hscode}\SaveRestoreHook
\column{B}{@{}>{\hspre}l<{\hspost}@{}}%
\column{E}{@{}>{\hspre}l<{\hspost}@{}}%
\>[B]{}\Varid{getData}\mathbin{::}[\mskip1.5mu (\Conid{User},[\mskip1.5mu (\Conid{Key},\Conid{Data})\mskip1.5mu])\mskip1.5mu]\to \Conid{User}\to \Conid{Key}\to \Conid{Maybe}\;\Conid{Data}{}\<[E]%
\\
\>[B]{}\Varid{getData}\;\Varid{db}\;\Varid{u}\;\Varid{k}\mathrel{=}\Varid{lookup}\;\Varid{u}\;\Varid{db}\bind \Varid{lookup}\;\Varid{k}{}\<[E]%
\ColumnHook
\end{hscode}\resethooks
If a user~\ensuremath{\Varid{u}} has an entry~\ensuremath{\Varid{d}} associated with a key~\ensuremath{\Varid{k}} in the database, \ensuremath{\Varid{getData}\;\Varid{u}\;\Varid{k}} returns \ensuremath{\Conid{Just}\;\Varid{d}}; otherwise, it returns \ensuremath{\Conid{Nothing}}. In the latter case we might want to inform the user `why' the program is not able to deliver data: they might have misspelled their username, which means that \ensuremath{\Varid{lookup}\;\Varid{u}} will fail, or they might have tried to read from a missing key, which means that \ensuremath{\Varid{lookup}\;\Varid{k}} will fail. Unfortunately, the \ensuremath{\Conid{Maybe}} monad does not allow one to observe at what point a failing program actually fails. We need to structure our function using a more sophisticated monad.

What are the desired properties of such a monad? For sure, we want it to employ the same kind of effects as \ensuremath{\Conid{Maybe}}, so that we do not have to alter the logic of our program. We would like to have a \ensuremath{\Varid{lift}} operation, which allows us to automatically translate some operations from \ensuremath{\Conid{Maybe}} into the new monad, so that it is not necessary to rewrite standard functions like \ensuremath{\Varid{lookup}}. But most importantly, it must also reveal some observations about the course of execution (a \emph{trace}), such as the number of successful subcomputations, from which we can extract the desired information about the point of failure. One possibility is to explicitly accumulate such observations inside the computation, using monads like \ensuremath{\Conid{Writer}} or \ensuremath{\Conid{Exception}}. In this article, however, we take a different approach: we automate the accumulation inside the monadic structure. The accumulation is transparent inside the computation; that is, it cannot affect the course of execution, and is revealed only on the outside.

We aim for maximum genericity: we construct transformers that add traces to arbitrary monads. Our main tools are free monads (Section~\ref{sec:free}), which have the ability to represent the very structure of monadic computations. They provide, in a sense, the most informative traces possible. Then, we introduce a transformer, called \ensuremath{\mathit{Nest}}, which allows one to mix the free and effectful computations provided by a monad (Section~\ref{sec:self}). The genericity pays off, and we find a number of applications for tracing monads:

\begin{itemize}
\item Traces of computations can be interpreted in different ways. In Section~\ref{sec:cut} we show an example in which, by revealing the inner structure of a computation in the \ensuremath{\Conid{List}} monad, we can define the Prolog \ensuremath{\Varid{cut}} operator.

\item Computations encapsulated in monadic expressions are often monolithic. They are supposed to produce final values only, so there is little space for non-termination. But, assuming non-strict semantics, we can see computations as entities that lazily unfold a trace. So, even if a computation is non-terminating, we can benefit from its infinite trace. If we interpret free parts of a \ensuremath{\mathit{Nest}} value as terms, we discover that \ensuremath{\mathit{Nest}} is a generalisation of Capretta's partiality monad (see Section~\ref{sec:partiality}).

\item For the same reasons, traces are a useful tool for modelling impure interaction with the environment. In Section~\ref{sec:future}, we sketch out some future work on a novel, coinductive approach to the functional semantics of effects.
\end{itemize}

\section{Simple tracing with free monads}\label{sec:free}

Moggi~\cite{Moggi:1991:NCM:116981.116984} called monads \emph{notions of computation}, because they describe computational effects in a way that abstracts from the type of values produced by a computation. In a sense, from the categorical point of view, monads abstract even from the exact values produced by the computation, since \ensuremath{\Varid{return}} and \ensuremath{\Varid{join}} are natural transformations. It is the \ensuremath{\Varid{bind}} operator (defined as \ensuremath{\Varid{m}\bind \Varid{f}\mathrel{=}\Varid{join}\;(\Varid{fmap}\;\Varid{f}\;\Varid{m})}) that mixes the functorial (looking only at the values) \ensuremath{\Varid{fmap}} and parametric (looking only at the structure) \ensuremath{\Varid{join}}. The laws for monads and functors entail the following equality, called \emph{naturality} of join.
\begin{hscode}\SaveRestoreHook
\column{B}{@{}>{\hspre}l<{\hspost}@{}}%
\column{E}{@{}>{\hspre}l<{\hspost}@{}}%
\>[B]{}\Varid{fmap}\;\Varid{f}\mathbin{\circ}\Varid{join}\equiv \Varid{join}\mathbin{\circ}\Varid{fmap}\;(\Varid{fmap}\;\Varid{f}){}\<[E]%
\ColumnHook
\end{hscode}\resethooks
Read as a transformation from left to right, it allows one to move occurrences of \ensuremath{\Varid{join}} to the left of a composition, and occurrences of \ensuremath{\Varid{fmap}} to the right. This way, we can split computations with multiple steps into a ``mapping'' part and a ``joining'' part. For example, consider a computation \ensuremath{\Varid{m}\bind \Varid{f}\bind \Varid{g}}. It can be split as follows.
\begin{hscode}\SaveRestoreHook
\column{B}{@{}>{\hspre}l<{\hspost}@{}}%
\column{56}{@{}>{\hspre}l<{\hspost}@{}}%
\column{E}{@{}>{\hspre}l<{\hspost}@{}}%
\>[B]{}\Varid{m}\bind \Varid{f}\bind \Varid{g}\equiv (\Varid{join}\mathbin{\circ}\Varid{fmap}\;\Varid{g}\mathbin{\circ}\Varid{join}\mathbin{\circ}\Varid{fmap}\;\Varid{f})\;\Varid{m}\equiv {}\<[56]%
\>[56]{}(\Varid{join}\mathbin{\circ}\Varid{join}\mathbin{\circ}\Varid{fmap}\;(\Varid{fmap}\;\Varid{g})\mathbin{\circ}\Varid{fmap}\;\Varid{f})\;\Varid{m}{}\<[E]%
\ColumnHook
\end{hscode}\resethooks
Intuitively, we can see the ``mapping'' part (that is, \ensuremath{\Varid{fmap}\;(\Varid{fmap}\;\Varid{g})\mathbin{\circ}\Varid{fmap}\;\Varid{f}}) as the construction of the computation and the ``joining'' part (\ensuremath{\Varid{join}\mathbin{\circ}\Varid{join}}) as the execution of effects.

Our first approach to tracing is to capture the mapping part of a computation. We suspend execution of the \ensuremath{\Varid{join}} operators of the monad, so that we can isolate and examine the elements from which the computation is composed.

\subsection{Free monads}

The type of the mapping part is different for different numbers of nested occurrences of the \ensuremath{\Varid{fmap}} function, and so for computations consisting of different numbers of steps. For example,
\begin{hscode}\SaveRestoreHook
\column{B}{@{}>{\hspre}l<{\hspost}@{}}%
\column{45}{@{}>{\hspre}l<{\hspost}@{}}%
\column{E}{@{}>{\hspre}l<{\hspost}@{}}%
\>[B]{}\Conid{Just}\;\text{\tt 'a'}{}\<[45]%
\>[45]{}\mathbin{::}\Conid{Maybe}\;\Conid{Char}{}\<[E]%
\\
\>[B]{}\Varid{fmap}\;\Conid{Just}\;(\Conid{Just}\;\text{\tt 'a'}){}\<[45]%
\>[45]{}\mathbin{::}\Conid{Maybe}\;(\Conid{Maybe}\;\Conid{Char}){}\<[E]%
\\
\>[B]{}\Varid{fmap}\;(\Varid{fmap}\;\Conid{Just})\;(\Varid{fmap}\;\Conid{Just}\;(\Conid{Just}\;\text{\tt 'a'}))\ {}\<[45]%
\>[45]{}\mathbin{::}\Conid{Maybe}\;(\Conid{Maybe}\;(\Conid{Maybe}\;\Conid{Char})){}\<[E]%
\ColumnHook
\end{hscode}\resethooks
A reasonable class of datatypes in which to store such expressions are the \emph{free monads}, also known as \emph{\ensuremath{\Varid{f}}-generated trees}. Allowing ourselves to define monads in terms of \ensuremath{\Varid{fmap}}, \ensuremath{\Varid{return}}, and \ensuremath{\Varid{join}}, rather than the \ensuremath{\Varid{return}} and \ensuremath{\bind } required by Haskell, we can write:
\begin{hscode}\SaveRestoreHook
\column{B}{@{}>{\hspre}l<{\hspost}@{}}%
\column{3}{@{}>{\hspre}l<{\hspost}@{}}%
\column{22}{@{}>{\hspre}l<{\hspost}@{}}%
\column{30}{@{}>{\hspre}l<{\hspost}@{}}%
\column{32}{@{}>{\hspre}l<{\hspost}@{}}%
\column{E}{@{}>{\hspre}l<{\hspost}@{}}%
\>[B]{}\mathbf{data}\;\Conid{Free}\;\Varid{f}\;\Varid{a}\mathrel{=}\Conid{Wrap}\;(\Varid{f}\;(\Conid{Free}\;\Varid{f}\;\Varid{a}))\mid \Conid{Return}\;\Varid{a}{}\<[E]%
\\[\blanklineskip]%
\>[B]{}\mathbf{instance}\;\Conid{Functor}\;\Varid{f}\Rightarrow \Conid{Functor}\;(\Conid{Free}\;\Varid{f})\;\mathbf{where}{}\<[E]%
\\
\>[B]{}\hsindent{3}{}\<[3]%
\>[3]{}\Varid{fmap}\;\Varid{g}\;(\Conid{Return}\;\Varid{a}){}\<[22]%
\>[22]{}\mathrel{=}\Conid{Return}\;{}\<[32]%
\>[32]{}(\Varid{g}\;\Varid{a}){}\<[E]%
\\
\>[B]{}\hsindent{3}{}\<[3]%
\>[3]{}\Varid{fmap}\;\Varid{g}\;(\Conid{Wrap}\;\Varid{f}){}\<[22]%
\>[22]{}\mathrel{=}\Conid{Wrap}\;{}\<[30]%
\>[30]{}(\Varid{fmap}\;(\Varid{fmap}\;\Varid{g})\;\Varid{f}){}\<[E]%
\\[\blanklineskip]%
\>[B]{}\mathbf{instance}\;\Conid{Functor}\;\Varid{f}\Rightarrow \Conid{Monad}\;(\Conid{Free}\;\Varid{f})\;\mathbf{where}{}\<[E]%
\\
\>[B]{}\hsindent{3}{}\<[3]%
\>[3]{}\Varid{return}{}\<[22]%
\>[22]{}\mathrel{=}\Conid{Return}{}\<[E]%
\\[\blanklineskip]%
\>[B]{}\hsindent{3}{}\<[3]%
\>[3]{}\Varid{join}\;(\Conid{Return}\;\Varid{f}){}\<[22]%
\>[22]{}\mathrel{=}\Varid{f}{}\<[E]%
\\
\>[B]{}\hsindent{3}{}\<[3]%
\>[3]{}\Varid{join}\;(\Conid{Wrap}\;\Varid{f}){}\<[22]%
\>[22]{}\mathrel{=}\Conid{Wrap}\;(\Varid{fmap}\;\Varid{join}\;\Varid{f}){}\<[E]%
\ColumnHook
\end{hscode}\resethooks
Each level of the type constructor of a monad corresponds to a level in the \ensuremath{\Conid{Free}} data structure. We store \ensuremath{\Conid{Just}\;\text{\tt 'a'}\mathbin{::}\Conid{Maybe}\;\Conid{Char}} as \ensuremath{\Conid{Wrap}\;(\Conid{Just}\;(\Conid{Return}\;\text{\tt 'a'}))}, and \ensuremath{\Conid{Just}\;(\Conid{Just}\;(\text{\tt 'a'}))\mathbin{::}\Conid{Maybe}\;(\Conid{Maybe}\;\Conid{Char})} as \ensuremath{\Conid{Wrap}\;(\Conid{Just}\;(\Conid{Wrap}\;(\Conid{Just}\;(\Conid{Return}\;\text{\tt 'a'}))))}, and so on.

\subsection{Monad transformers with \ensuremath{\Varid{drop}}}\label{sec:monadTrans}

The type \ensuremath{\Conid{Free}\;\Varid{m}\;\Varid{a}} can be seen as a datatype of terms generated by the signature \ensuremath{\Varid{m}} (a functor) and a set of variables~\ensuremath{\Varid{a}}. Even if \ensuremath{\Varid{m}} is a monad, \ensuremath{\Conid{Free}\;\Varid{m}} cannot depend on any effects provided by \ensuremath{\Varid{m}}; the join operation for \ensuremath{\Conid{Free}\;\Varid{m}} performs only substitution, and is independent of the join for \ensuremath{\Varid{m}}.

In order to couple a monad \ensuremath{\Varid{m}} and the \ensuremath{\Varid{m}}-generated free monad, we need the notion of \emph{monad transformers}~\cite{Liang:1995:MTM:199448.199528}. A monad transformer with \ensuremath{\Varid{drop}} is a relation between two monads, \ensuremath{\Varid{m}} and \ensuremath{\Varid{t}}, characterised by two functions, \ensuremath{\Varid{lift}} and \ensuremath{\Varid{drop}}, which translate computations in \ensuremath{\Varid{m}} into computations in \ensuremath{\Varid{t}}, and \emph{vice versa}. The relationship can be defined as the following two-parameter Haskell type class. Though in functional programming \ensuremath{\Varid{drop}} is rarely considered to be a member of \ensuremath{\Conid{MonadTrans}}, it plays an important role here. For the moment, we forget that we usually insist that monad transformers are subject to a certain set of algebraic laws.
\begin{hscode}\SaveRestoreHook
\column{B}{@{}>{\hspre}l<{\hspost}@{}}%
\column{3}{@{}>{\hspre}l<{\hspost}@{}}%
\column{9}{@{}>{\hspre}l<{\hspost}@{}}%
\column{17}{@{}>{\hspre}l<{\hspost}@{}}%
\column{E}{@{}>{\hspre}l<{\hspost}@{}}%
\>[B]{}\mathbf{class}\;(\Conid{Monad}\;\Varid{m},\ \Conid{Monad}\;\Varid{t})\Rightarrow \Conid{MonadTrans}\;\Varid{m}\;\Varid{t}\mid \Varid{t}\to \Varid{m}\;\mathbf{where}{}\<[E]%
\\
\>[B]{}\hsindent{3}{}\<[3]%
\>[3]{}\Varid{lift}{}\<[9]%
\>[9]{}\mathbin{::}\Varid{m}\;\Varid{a}{}\<[17]%
\>[17]{}\to \Varid{t}\;\Varid{a}{}\<[E]%
\\
\>[B]{}\hsindent{3}{}\<[3]%
\>[3]{}\Varid{drop}{}\<[9]%
\>[9]{}\mathbin{::}\Varid{t}\;\Varid{a}{}\<[17]%
\>[17]{}\to \Varid{m}\;\Varid{a}{}\<[E]%
\ColumnHook
\end{hscode}\resethooks
For any monad \ensuremath{\Varid{m}}, we define an instance of \ensuremath{\Conid{MonadTrans}\;\Varid{m}\;(\Conid{Free}\;\Varid{m})} as follows. The \ensuremath{\Varid{lift}} operation is straightforward, as it only wraps the value and maps the \ensuremath{\Conid{Return}} constructor. The \ensuremath{\Varid{drop}} operation traverses the structure and flattens each level, thus performing suspended \ensuremath{\Varid{bind}}s of \ensuremath{\Varid{m}}.
\begin{hscode}\SaveRestoreHook
\column{B}{@{}>{\hspre}l<{\hspost}@{}}%
\column{3}{@{}>{\hspre}l<{\hspost}@{}}%
\column{20}{@{}>{\hspre}l<{\hspost}@{}}%
\column{E}{@{}>{\hspre}l<{\hspost}@{}}%
\>[B]{}\mathbf{instance}\;(\Conid{Functor}\;\Varid{m},\ \Conid{Monad}\;\Varid{m})\Rightarrow \Conid{MonadTrans}\;\Varid{m}\;(\Conid{Free}\;\Varid{m})\;\mathbf{where}{}\<[E]%
\\
\>[B]{}\hsindent{3}{}\<[3]%
\>[3]{}\Varid{lift}{}\<[20]%
\>[20]{}\mathrel{=}\Conid{Wrap}\mathbin{\circ}\Varid{fmap}\;\Conid{Return}{}\<[E]%
\\[\blanklineskip]%
\>[B]{}\hsindent{3}{}\<[3]%
\>[3]{}\Varid{drop}\;(\Conid{Return}\;\Varid{a}){}\<[20]%
\>[20]{}\mathrel{=}\Varid{return}\;\Varid{a}{}\<[E]%
\\
\>[B]{}\hsindent{3}{}\<[3]%
\>[3]{}\Varid{drop}\;(\Conid{Wrap}\;\Varid{m}){}\<[20]%
\>[20]{}\mathrel{=}\Varid{m}\bind \Varid{drop}{}\<[E]%
\ColumnHook
\end{hscode}\resethooks

\subsection{Examples}

Now, we can test our tracing free monad transformer on the \ensuremath{\Conid{Maybe}} monad. The \ensuremath{\Varid{lift}} function allows us to translate any computation in \ensuremath{\Varid{m}} into a computation in \ensuremath{\Conid{Free}\;\Varid{m}}. To get the original, non-traced computation back, we use the \ensuremath{\Varid{drop}} function.

We can see every composition \ensuremath{\Conid{Wrap}\mathbin{\circ}\Conid{Just}} as a ``tick'', given for each lifted successful subcomputation. The trace forms a unary counter, storing the number of ticks. The final value of the computation is stored in the last \ensuremath{\Conid{Wrap}}. Consider the following conversation with the Haskell interactive shell.
\begin{hscode}\SaveRestoreHook
\column{B}{@{}>{\hspre}l<{\hspost}@{}}%
\column{E}{@{}>{\hspre}l<{\hspost}@{}}%
\>[B]{}\mathtt{\triangleright}\;\mathbf{do}\;\{\mskip1.5mu \Varid{lift}\;(\Conid{Just}\;\mathrm{2});\ \Varid{lift}\;(\Conid{Just}\;\mathrm{4});\ \Varid{lift}\;\Conid{Nothing}\mskip1.5mu\}{}\<[E]%
\\
\>[B]{}\Conid{Wrap}\;(\Conid{Just}\;(\Conid{Wrap}\;(\Conid{Just}\;(\Conid{Wrap}\;\Conid{Nothing})))){}\<[E]%
\\[\blanklineskip]%
\>[B]{}\mathtt{\triangleright}\;\Varid{drop}\;(\mathbf{do}\;\{\mskip1.5mu \Varid{lift}\;(\Conid{Just}\;\mathrm{2});\ \Varid{lift}\;(\Conid{Just}\;\mathrm{4});\ \Varid{lift}\;\Conid{Nothing}\mskip1.5mu\}){}\<[E]%
\\
\>[B]{}\Conid{Nothing}{}\<[E]%
\ColumnHook
\end{hscode}\resethooks
Similarly, for the \ensuremath{\Conid{Writer}} monad, we can get a list of all the values appended to the monoid followed by the final return value. (For brevity, we show \ensuremath{\Conid{Writer}\;(\Varid{a},\Conid{Sum}\;\Varid{s})} as \ensuremath{(\Varid{s},\Varid{a})}.)
\begin{hscode}\SaveRestoreHook
\column{B}{@{}>{\hspre}l<{\hspost}@{}}%
\column{E}{@{}>{\hspre}l<{\hspost}@{}}%
\>[B]{}\mathtt{\triangleright}\;\mathbf{do}\;\{\mskip1.5mu \Varid{lift}\;(\Varid{tell}\;(\Conid{Sum}\;\mathrm{2}));\ \Varid{lift}\;(\Varid{tell}\;(\Conid{Sum}\;\mathrm{3}));\ \Varid{lift}\;(\Varid{tell}\;(\Conid{Sum}\;\mathrm{7}));\ \Varid{return}\;\text{\tt 'a'}\mskip1.5mu\}{}\<[E]%
\\
\>[B]{}\Conid{Wrap}\;(\mathrm{2},\Conid{Wrap}\;(\mathrm{3},\Conid{Wrap}\;(\mathrm{7},\Conid{Return}\;\text{\tt 'a'}))){}\<[E]%
\\[\blanklineskip]%
\>[B]{}\mathtt{\triangleright}\;\Varid{drop}\;(\mathbf{do}\;\{\mskip1.5mu \Varid{lift}\;(\Varid{tell}\;(\Conid{Sum}\;\mathrm{2}));\ \Varid{lift}\;(\Varid{tell}\;(\Conid{Sum}\;\mathrm{3}));\ \Varid{lift}\;(\Varid{tell}\;(\Conid{Sum}\;\mathrm{7}));\ \Varid{return}\;\text{\tt 'a'}\mskip1.5mu\}){}\<[E]%
\\
\>[B]{}(\mathrm{12},\text{\tt 'a'}){}\<[E]%
\ColumnHook
\end{hscode}\resethooks
An important thing to notice is that non-terminating computations in \ensuremath{\Conid{Writer}} do not make much sense. For example, the following computation just diverges.
\begin{hscode}\SaveRestoreHook
\column{B}{@{}>{\hspre}l<{\hspost}@{}}%
\column{E}{@{}>{\hspre}l<{\hspost}@{}}%
\>[B]{}\mathtt{\triangleright}\;\mathbf{let}\;\Varid{w}\;\Varid{n}\mathrel{=}\mathbf{do}\;\{\mskip1.5mu \Varid{tell}\;(\Conid{Sum}\;\Varid{n});\ \Varid{w}\;(\Varid{n}\mathbin{+}\mathrm{1})\mskip1.5mu\}\;\mathbf{in}\;\Varid{w}\;\mathrm{0}{}\<[E]%
\\
\>[B]{}(\ \!\mathord{\ast}\mathord{\ast}\mathord{\ast}\;\ \mathrm{Exception}\mathbin{:}\ \mathrm{stack}\;\mathrm{overflow}{}\<[E]%
\ColumnHook
\end{hscode}\resethooks
In contrast, with the tracing version of \ensuremath{\Conid{Writer}}, we can enjoy an infinite stream of actions that happen during the execution:
\begin{hscode}\SaveRestoreHook
\column{B}{@{}>{\hspre}l<{\hspost}@{}}%
\column{E}{@{}>{\hspre}l<{\hspost}@{}}%
\>[B]{}\mathtt{\triangleright}\;\mathbf{let}\;\Varid{w}\;\Varid{n}\mathrel{=}\mathbf{do}\;\{\mskip1.5mu \Varid{lift}\;(\Varid{tell}\;(\Conid{Sum}\;\Varid{n}));\ \Varid{w}\;(\Varid{n}\mathbin{+}\mathrm{1})\mskip1.5mu\}\;\mathbf{in}\;\Varid{w}\;\mathrm{0}{}\<[E]%
\\
\>[B]{}\Conid{Wrap}\;(\mathrm{0},\Conid{Wrap}\;(\mathrm{1},\Conid{Wrap}\;(\mathrm{2},\Conid{Wrap}\;(\mathrm{3},\Conid{Wrap}\;(\mathrm{4},\Conid{Wrap}\;(\mathrm{5},\Conid{Wrap}\;(\mathrm{6},\Conid{Wrap}\;(\mathrm{7},\Conid{Wrap}\;(\mathrm{8},\Conid{Wrap}.\mathinner{\ldotp\ldotp}{}\<[E]%
\ColumnHook
\end{hscode}\resethooks

\section{More advanced tracing with free structures}\label{sec:self}
\def\commentbegin{\quad\{\ }
\def\commentend{\}}

Tracing computations with free structures is not very flexible: every bind and every join is suspended, creating a new level of structure every time a monadic action is called. In some circumstances we would like to trace only selected parts of the computation---perhaps we are confident that the other parts always succeed, or we want to treat selected pieces of the computation monolithically, and we are not interested in a fine-grained report about their behaviour.

Another issue is the algebraic properties of monad transformers with \ensuremath{\Varid{drop}}. Intuitively, a pair of monads \ensuremath{\Varid{m}} and \ensuremath{\Varid{t}} are related as a monad transformer if \ensuremath{\Varid{t}} incorporates at least the same effects as \ensuremath{\Varid{m}}. This is usually formalised with the following equalities~\cite{Hinze:2000:DBM:351240.351258}.
%
\begin{hscode}\SaveRestoreHook
\column{B}{@{}>{\hspre}l<{\hspost}@{}}%
\column{22}{@{}>{\hspre}l<{\hspost}@{}}%
\column{E}{@{}>{\hspre}l<{\hspost}@{}}%
\>[B]{}\Varid{lift}\;(\Varid{return}\;\Varid{a}){}\<[22]%
\>[22]{}\mathrel{=}\Varid{return}\;\Varid{a}{}\<[E]%
\\
\>[B]{}\Varid{lift}\;(\Varid{m}\bind \Varid{f}){}\<[22]%
\>[22]{}\mathrel{=}\Varid{lift}\;\Varid{m}\bind (\Varid{lift}\mathbin{\circ}\Varid{f}){}\<[E]%
\\[\blanklineskip]%
\>[B]{}\Varid{drop}\;(\Varid{return}\;\Varid{a}){}\<[22]%
\>[22]{}\mathrel{=}\Varid{return}\;\Varid{a}{}\<[E]%
\\
\>[B]{}\Varid{drop}\;(\Varid{lift}\;\Varid{m}\bind \Varid{f}){}\<[22]%
\>[22]{}\mathrel{=}\Varid{m}\bind \Varid{drop}\mathbin{\circ}\Varid{f}{}\<[E]%
\ColumnHook
\end{hscode}\resethooks
%
%
The first two equalities state that \ensuremath{\Varid{lift}} is a monad morphism. The instance \ensuremath{\Conid{MonadTrans}\;\Varid{m}\;(\Conid{Free}\;\Varid{m})} from Section~\ref{sec:monadTrans} does not have this property. For example, \ensuremath{\Varid{lift}\;(\Conid{Just}\;\mathrm{1})\sequ \Varid{lift}\;(\Conid{Just}\;\mathrm{2})} is equal to \ensuremath{\Conid{Wrap}\;(\Conid{Just}\;(\Conid{Wrap}\;(\Conid{Just}\;(\Conid{Return}\;\mathrm{2}))))}, while \ensuremath{\Varid{lift}\;(\Conid{Just}\;\mathrm{1}\sequ \Conid{Just}\;\mathrm{2})} is equal to \ensuremath{\Conid{Wrap}\;(\Conid{Just}\;(\Conid{Return}\;\mathrm{2}))}.

For these two reasons, we abandon the idea of using free monads directly to trace computations. Instead, in this section we introduce a general class, \ensuremath{\Conid{MonadTrace}}, which allows one to specify the points at which to make observations about the execution, and a corresponding version of the \ensuremath{\Conid{Free}} monad that is a proper monad transformer.

\subsection{The \ensuremath{\Conid{MonadTrace}} class}\label{sec:tracer}

The \ensuremath{\Conid{MonadTrace}} class introduces a single monadic value, \ensuremath{\Varid{mark}}. Intuitively, this is an operation that stores the current state of execution in the trace.
\begin{hscode}\SaveRestoreHook
\column{B}{@{}>{\hspre}l<{\hspost}@{}}%
\column{3}{@{}>{\hspre}l<{\hspost}@{}}%
\column{E}{@{}>{\hspre}l<{\hspost}@{}}%
\>[B]{}\mathbf{class}\;\Conid{Monad}\;\Varid{t}\Rightarrow \Conid{MonadTrace}\;\Varid{t}\;\mathbf{where}{}\<[E]%
\\
\>[B]{}\hsindent{3}{}\<[3]%
\>[3]{}\Varid{mark}\mathbin{::}\Varid{t}\;(){}\<[E]%
\ColumnHook
\end{hscode}\resethooks
We call a monad \ensuremath{\Varid{v}} a \emph{tracer} of a monad \ensuremath{\Varid{m}}, if \ensuremath{\Varid{m}} and \ensuremath{\Varid{v}} form a monad transformer (\ensuremath{\Conid{MonadTrans}\;\Varid{m}\;\Varid{v}}), and \ensuremath{\Varid{v}} is an instance of \ensuremath{\Conid{MonadTrace}}.
%
Additionally, \ensuremath{\Varid{lift}}, \ensuremath{\Varid{drop}} and \ensuremath{\Varid{mark}} should satisfy the following equalities.
\savecolumns
\begin{hscode}\SaveRestoreHook
\column{B}{@{}>{\hspre}l<{\hspost}@{}}%
\column{7}{@{}>{\hspre}l<{\hspost}@{}}%
\column{19}{@{}>{\hspre}l<{\hspost}@{}}%
\column{E}{@{}>{\hspre}l<{\hspost}@{}}%
\>[B]{}\Varid{lift}\;{}\<[7]%
\>[7]{}(\Varid{return}\;\Varid{a}){}\<[19]%
\>[19]{}\mathrel{=}\Varid{return}\;\Varid{a}{}\<[E]%
\\
\>[B]{}\Varid{lift}\;{}\<[7]%
\>[7]{}(\Varid{m}\bind \Varid{f}){}\<[19]%
\>[19]{}\mathrel{=}\Varid{lift}\;\Varid{m}\bind (\Varid{lift}\mathbin{\circ}\Varid{f}){}\<[E]%
\\[\blanklineskip]%
\>[B]{}\Varid{drop}\;{}\<[7]%
\>[7]{}(\Varid{return}\;\Varid{a}){}\<[19]%
\>[19]{}\mathrel{=}\Varid{return}\;\Varid{a}{}\<[E]%
\\
\>[B]{}\Varid{drop}\;{}\<[7]%
\>[7]{}(\Varid{v}\bind \Varid{g}){}\<[19]%
\>[19]{}\mathrel{=}\Varid{drop}\;\Varid{v}\bind (\Varid{drop}\mathbin{\circ}\Varid{g}){}\<[E]%
\\[\blanklineskip]%
\>[B]{}\Varid{drop}\;{}\<[7]%
\>[7]{}\Varid{mark}{}\<[19]%
\>[19]{}\mathrel{=}\Varid{return}\;(){}\<[E]%
\ColumnHook
\end{hscode}\resethooks
The tracer \ensuremath{\Varid{v}} cannot perform more effects than the monad \ensuremath{\Varid{m}}, except for tracing with the \ensuremath{\Varid{mark}} operation. Nonetheless, tracing does not affect the course of computation in any way observable from inside of the \ensuremath{\Varid{v}}-computation, hence both \ensuremath{\Varid{lift}} and \ensuremath{\Varid{drop}} are monad morphisms. Note that the laws entail \ensuremath{\Varid{drop}\mathbin{\circ}\Varid{lift}\mathrel{=}\Varid{id}}.

We use the \ensuremath{\Varid{mark}} gadget wherever we want to make an observation. In
the following example, the intuitive semantics of a tracer for the
\ensuremath{\Conid{Maybe}} monad is that we get a tick whenever the computation is still
successful when placing a mark.
%
\begin{hscode}\SaveRestoreHook
\column{B}{@{}>{\hspre}l<{\hspost}@{}}%
\column{7}{@{}>{\hspre}l<{\hspost}@{}}%
\column{E}{@{}>{\hspre}l<{\hspost}@{}}%
\>[B]{}\mathbf{do}\;\{\mskip1.5mu {}\<[7]%
\>[7]{}\Varid{x}\leftarrow \Varid{lift}\;\Varid{m}_{0};\ \Varid{y}\leftarrow \Varid{lift}\;\Varid{m}_{1};\ \Varid{mark};\ \Varid{z}\leftarrow \Varid{lift}\;\Varid{m}_{2};\ \Varid{mark};\ \Varid{return}\;(\Varid{x}\mathbin{+}\Varid{y}\mathbin{+}\Varid{z})\mskip1.5mu\}{}\<[E]%
\ColumnHook
\end{hscode}\resethooks
That means that if \ensuremath{\Varid{m}_{0}} is successful, but \ensuremath{\Varid{m}_{1}} fails, no ticks are
stored in the trace (intuitively, it is equivalent to \ensuremath{\Conid{Nothing}}). Only
if both \ensuremath{\Varid{m}_{0}} and \ensuremath{\Varid{m}_{1}} are successful, the \ensuremath{\Varid{mark}} operation stores this
success (the trace is of the form \ensuremath{\Conid{Just}\;\Varid{a}}, where \ensuremath{\Varid{a}} is a result of
the rest of the computation). If \ensuremath{\Varid{m}_{2}} is also successful, the second
call to \ensuremath{\Varid{mark}} stores this information in the trace (and the trace is
of the form \ensuremath{\Conid{Just}\;(\Conid{Just}\;\Varid{a})}, where \ensuremath{\Varid{a}\mathrel{=}\Varid{return}\;(\Varid{x}\mathbin{+}\Varid{y}\mathbin{+}\Varid{z})\mathrel{=}\Conid{Just}\;(\Varid{x}\mathbin{+}\Varid{y}\mathbin{+}\Varid{z})}).

We also define a convenient function, \ensuremath{\Varid{mind}}, to perform a traced
lifting.
\begin{hscode}\SaveRestoreHook
\column{B}{@{}>{\hspre}l<{\hspost}@{}}%
\column{30}{@{}>{\hspre}l<{\hspost}@{}}%
\column{E}{@{}>{\hspre}l<{\hspost}@{}}%
\>[B]{}\Varid{mind}\mathbin{::}(\Conid{MonadTrans}\;\Varid{m}\;\Varid{v},\ {}\<[30]%
\>[30]{}\Conid{MonadTrace}\;\Varid{v})\Rightarrow \Varid{m}\;\Varid{a}\to \Varid{v}\;\Varid{a}{}\<[E]%
\\
\>[B]{}\Varid{mind}\;\Varid{m}\mathrel{=}\mathbf{do}\;\{\mskip1.5mu \Varid{x}\leftarrow \Varid{lift}\;\Varid{m};\ \Varid{mark};\ \Varid{return}\;\Varid{x}\mskip1.5mu\}{}\<[E]%
\ColumnHook
\end{hscode}\resethooks
Then the previous example can be written more concisely:
\begin{hscode}\SaveRestoreHook
\column{B}{@{}>{\hspre}l<{\hspost}@{}}%
\column{E}{@{}>{\hspre}l<{\hspost}@{}}%
\>[B]{}\mathbf{do}\;\{\mskip1.5mu \Varid{x}\leftarrow \Varid{lift}\;\Varid{m}_{0};\ \Varid{y}\leftarrow \Varid{mind}\;\Varid{m}_{1};\ \Varid{z}\leftarrow \Varid{mind}\;\Varid{m}_{2};\ \Varid{return}\;(\Varid{x}\mathbin{+}\Varid{y}\mathbin{+}\Varid{z})\mskip1.5mu\}{}\<[E]%
\ColumnHook
\end{hscode}\resethooks

\subsection{The \ensuremath{\mathit{Nest}} monad} \label{sec:self-monad}

Free monads allow one to capture the structure of an \ensuremath{\Varid{m}}-computation as data; but for tracing, we need also to be able to perform some parts of the computations (the lifted ones) immediately. This suggests considering the composition of the two monads \ensuremath{\Varid{m}} and \ensuremath{\Conid{Free}\;\Varid{m}}, in one order or the other. In fact, because we want the lifted parts to be performed immediately, the appropriate order of composition is to have \ensuremath{\Varid{m}} on the outside and \ensuremath{\Conid{Free}\;\Varid{m}} inside. We therefore define:
\begin{hscode}\SaveRestoreHook
\column{B}{@{}>{\hspre}l<{\hspost}@{}}%
\column{3}{@{}>{\hspre}l<{\hspost}@{}}%
\column{E}{@{}>{\hspre}l<{\hspost}@{}}%
\>[B]{}\mathbf{newtype}\;\mathit{Nest}\;\Varid{m}\;\Varid{a}\mathrel{=}\mathit{Nest}\{\mskip1.5mu \mathit{unNest}\mathbin{::}\Varid{m}\;(\Conid{Free}\;\Varid{m}\;\Varid{a})\mskip1.5mu\}{}\<[E]%
\\[\blanklineskip]%
\>[B]{}\mathbf{instance}\;\Conid{Functor}\;\Varid{m}\Rightarrow \Conid{Functor}\;(\mathit{Nest}\;\Varid{m})\;\mathbf{where}{}\<[E]%
\\
\>[B]{}\hsindent{3}{}\<[3]%
\>[3]{}\Varid{fmap}\;\Varid{f}\;(\mathit{Nest}\;\Varid{m})\mathrel{=}\mathit{Nest}\;(\Varid{fmap}\;(\Varid{fmap}\;\Varid{f})\;\Varid{m}){}\<[E]%
\ColumnHook
\end{hscode}\resethooks
It remains to show that \ensuremath{\mathit{Nest}} can be given the structure of a
monad. We do this using Jones and Duponcheel's \ensuremath{\Varid{prod}}
construction~\cite{Jones93composingmonads}: given monad \ensuremath{\Conid{M}} with unit
\ensuremath{\Varid{return}_{\Conid{M}}} and multiplication \ensuremath{\Varid{join}_{\Conid{M}}}, and
similarly monad \ensuremath{\Conid{F}} with
\ensuremath{\Varid{return}_{\Conid{F}}} and \ensuremath{\Varid{join}_{\Conid{F}}}, the composition \ensuremath{\Conid{M}\;\Conid{F}} forms a monad
with unit and multiplication given by

\noindent
\begin{minipage}[l]{4cm}
\begin{hscode}\SaveRestoreHook
\column{B}{@{}>{\hspre}l<{\hspost}@{}}%
\column{9}{@{}>{\hspre}l<{\hspost}@{}}%
\column{E}{@{}>{\hspre}l<{\hspost}@{}}%
\>[B]{}\Varid{return}{}\<[9]%
\>[9]{}\mathrel{=}\Varid{return}_{\Conid{M}}\mathbin{\circ}\Varid{return}_{\Conid{F}}{}\<[E]%
\\
\>[B]{}\Varid{join}{}\<[9]%
\>[9]{}\mathrel{=}\Varid{join}_{\Conid{M}}\mathbin{\circ}\Varid{fmap}_{\Conid{M}}\;\Varid{prod}{}\<[E]%
\ColumnHook
\end{hscode}\resethooks
\end{minipage}
\hfill
%
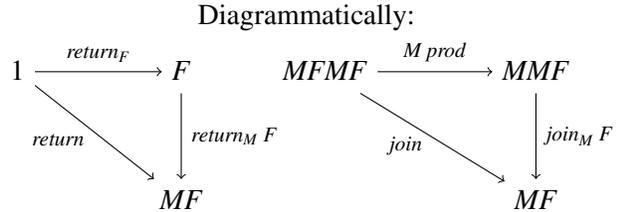
\begin{minipage}[l]{8.4cm}
\begin{center}
Diagrammatically:\\
\begin{tikzpicture}[description/.style={fill=white,inner sep=1pt}, rotate=90]
\matrix (m) [matrix of math nodes, row sep=3em,
column sep=4em, text height=1.5ex, text depth=0.25ex]
{
\ensuremath{\mathrm{1}} \pgfmatrixnextcell \ensuremath{\Conid{F}}
\\
\pgfmatrixnextcell \ensuremath{\Conid{MF}}
\\
};
\path[->,font=\scriptsize]
(m-1-1) edge[bend right=0] node[auto] {\ensuremath{\Varid{return}_{\Conid{F}}}} (m-1-2)
(m-1-2) edge[bend right=0] node[auto] {\ensuremath{\Varid{return}_{\Conid{M}}\;\Conid{F}}} (m-2-2)
(m-1-1) edge[bend right=0] node[left, yshift=-3pt] {\ensuremath{\Varid{return}}} (m-2-2)
;
\end{tikzpicture}
\hspace{-0.5cm}
\begin{tikzpicture}[description/.style={fill=white,inner sep=1pt}, rotate=90]
\matrix (m) [matrix of math nodes, row sep=3em,
column sep=4em, text height=1.5ex, text depth=0.25ex]
{
\ensuremath{\Conid{MFMF}} \pgfmatrixnextcell \ensuremath{\Conid{MMF}}
\\
\pgfmatrixnextcell \ensuremath{\Conid{MF}}
\\
};
\path[->,font=\scriptsize]
(m-1-1) edge[bend right=0] node[auto] {\ensuremath{\Conid{M}\;\Varid{prod}}} (m-1-2)
(m-1-2) edge[bend right=0] node[auto] {\ensuremath{\Varid{join}_{\Conid{M}}\;\Conid{F}}} (m-2-2)
(m-1-1) edge[bend right=0] node[left, yshift=-3pt] {\ensuremath{\Varid{join}}} (m-2-2)
;
\end{tikzpicture}
\end{center}
\end{minipage}
\vspace{1em}

\noindent
provided that natural transformation \ensuremath{\Varid{prod}\mathbin{::}\Conid{F}\;\Conid{M}\;\Conid{F}\stackrel{.}{\rightarrow}\Conid{M}\;\Conid{F}} satisfies the three properties
\begin{hscode}\SaveRestoreHook
\column{B}{@{}>{\hspre}l<{\hspost}@{}}%
\column{23}{@{}>{\hspre}l<{\hspost}@{}}%
\column{39}{@{}>{\hspre}l<{\hspost}@{}}%
\column{E}{@{}>{\hspre}l<{\hspost}@{}}%
\>[B]{}\Varid{prod}\mathbin{\circ}\Varid{return}_{\Conid{F}}{}\<[23]%
\>[23]{}\mathrel{=}\Varid{id}{}\<[39]%
\>[39]{}\ \ \ (\mathrm{1}){}\<[E]%
\\
\>[B]{}\Varid{prod}\mathbin{\circ}\Varid{fmap}_{\Conid{F}}\;\Varid{return}{}\<[23]%
\>[23]{}\mathrel{=}\Varid{return}_{\Conid{M}}{}\<[39]%
\>[39]{}\ \ \ (\mathrm{2}){}\<[E]%
\\
\>[B]{}\Varid{prod}\mathbin{\circ}\Varid{fmap}_{\Conid{F}}\;\Varid{join}{}\<[23]%
\>[23]{}\mathrel{=}\Varid{join}\mathbin{\circ}\Varid{prod}{}\<[39]%
\>[39]{}\ \ \ (\mathrm{3}){}\<[E]%
\ColumnHook
\end{hscode}\resethooks

\noindent
Diagrammatically:\\
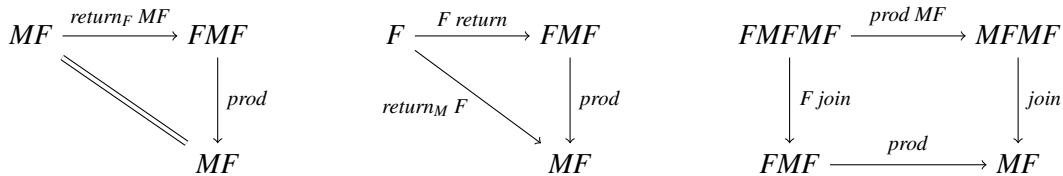
\begin{center}
\begin{minipage}[l]{14.7cm}
\begin{center}
\begin{tikzpicture}[description/.style={fill=white,inner sep=1pt}, rotate=90]
\matrix (m) [matrix of math nodes, row sep=3em,
column sep=4em, text height=1.5ex, text depth=0.25ex]
{
\ensuremath{\Conid{MF}} \pgfmatrixnextcell \ensuremath{\Conid{FMF}}
\\
\pgfmatrixnextcell \ensuremath{\Conid{MF}}
\\
};
\path[->,font=\scriptsize]
(m-1-1) edge[bend right=0] node[auto] {\ensuremath{\Varid{return}_{\Conid{F}}\;\Conid{MF}}} (m-1-2)
(m-1-2) edge[bend right=0] node[auto] {\ensuremath{\Varid{prod}}} (m-2-2);
\draw[double distance = 1.5pt] (m-1-1) -- (m-2-2);
\end{tikzpicture}
\hspace{1cm}
\begin{tikzpicture}[description/.style={fill=white,inner sep=1pt}, rotate=90]
\matrix (m) [matrix of math nodes, row sep=3em,
column sep=4em, text height=1.5ex, text depth=0.25ex]
{
\ensuremath{\Conid{F}} \pgfmatrixnextcell \ensuremath{\Conid{FMF}}
\\
\pgfmatrixnextcell \ensuremath{\Conid{MF}}
\\
};
\path[->,font=\scriptsize]
(m-1-1) edge[bend right=0] node[auto] {\ensuremath{\Conid{F}\;\Varid{return}}} (m-1-2)
(m-1-2) edge[bend right=0] node[auto] {\ensuremath{\Varid{prod}}} (m-2-2)
(m-1-1) edge[bend right=0] node[left, yshift=-4pt] {\ensuremath{\Varid{return}_{\Conid{M}}\;\Conid{F}}} (m-2-2)
;
\end{tikzpicture}
\hspace{1cm}
\begin{tikzpicture}[description/.style={fill=white,inner sep=1pt}, rotate=90]
\matrix (m) [matrix of math nodes, row sep=3em,
column sep=4em, text height=1.5ex, text depth=0.25ex]
{
\ensuremath{\Conid{FMFMF}} \pgfmatrixnextcell \ensuremath{\Conid{MFMF}}
\\
\ensuremath{\Conid{FMF}} \pgfmatrixnextcell \ensuremath{\Conid{MF}}
\\
};
\path[->,font=\scriptsize]
(m-1-1) edge[bend right=0] node[auto] {\ensuremath{\Varid{prod}\;\Conid{MF}}} (m-1-2)
(m-1-2) edge[bend right=0] node[auto] {\ensuremath{\Varid{join}}} (m-2-2)
(m-1-1) edge[bend right=0] node[auto] {\ensuremath{\Conid{F}\;\Varid{join}}} (m-2-1)
(m-2-1) edge[bend right=0] node[auto] {\ensuremath{\Varid{prod}}} (m-2-2)
;
\end{tikzpicture}
\end{center}
\end{minipage}
\end{center}
\vspace{0.5em}

\noindent
(In fact, the multiplication \ensuremath{\Varid{join}_{\Conid{F}}} of \ensuremath{\Conid{F}} is not used; all that is required of \ensuremath{\Conid{F}} is for it to be a \emph{premonad}.)
It turns out that the definition of \ensuremath{\Varid{prod}} is---if not quite forced
then at least---very strongly suggested by the necessity of satisfying
these three properties.

In our case, \ensuremath{\Conid{M}} is the monad \ensuremath{\Varid{m}}, and \ensuremath{\Conid{F}} is the datatype \ensuremath{\Conid{Free}\;\Varid{m}} from Section~\ref{sec:free}. For brevity, let us write \ensuremath{\Conid{M}} for \ensuremath{\Varid{fmap}_{\Conid{M}}}, and similarly \ensuremath{\Conid{F}} for \ensuremath{\Varid{fmap}_{\Conid{F}}}; let us also write the coproduct type former as \ensuremath{\mathbin{+}} and the coproduct morphism as \ensuremath{\mathbin{\triangledown}}, so that we can express the two constructors as one composite,
\begin{hscode}\SaveRestoreHook
\column{B}{@{}>{\hspre}l<{\hspost}@{}}%
\column{E}{@{}>{\hspre}l<{\hspost}@{}}%
\>[B]{}\Conid{Wrap}\mathbin{\triangledown}\Conid{Return}\mathbin{::}\Conid{M}\;\Conid{F}\mathbin{+}\mathrm{1}\stackrel{.}{\rightarrow}\Conid{F}{}\<[E]%
\ColumnHook
\end{hscode}\resethooks
and name its inverse,
\begin{hscode}\SaveRestoreHook
\column{B}{@{}>{\hspre}l<{\hspost}@{}}%
\column{E}{@{}>{\hspre}l<{\hspost}@{}}%
\>[B]{}\Varid{out}_{\Conid{F}}\mathbin{::}\Conid{F}\stackrel{.}{\rightarrow}\Conid{M}\;\Conid{F}\mathbin{+}\mathrm{1}{}\<[E]%
\ColumnHook
\end{hscode}\resethooks
Recall that the unit of the monad \ensuremath{\Conid{Free}\;\Varid{m}} is the constructor \ensuremath{\Conid{Return}}:
\begin{hscode}\SaveRestoreHook
\column{B}{@{}>{\hspre}l<{\hspost}@{}}%
\column{E}{@{}>{\hspre}l<{\hspost}@{}}%
\>[B]{}\Varid{return}_{\Conid{F}}\mathrel{=}\Conid{Return}{}\<[E]%
\ColumnHook
\end{hscode}\resethooks
Without loss of generality, we let
\begin{hscode}\SaveRestoreHook
\column{B}{@{}>{\hspre}l<{\hspost}@{}}%
\column{E}{@{}>{\hspre}l<{\hspost}@{}}%
\>[B]{}\Varid{prod}\mathbin{\circ}(\Conid{Wrap}\mathbin{\triangledown}\Conid{Return})\mathrel{=}\Varid{prod}_{1}\mathbin{\triangledown}\Varid{prod}_{2}{}\<[E]%
\ColumnHook
\end{hscode}\resethooks
so that
\begin{hscode}\SaveRestoreHook
\column{B}{@{}>{\hspre}l<{\hspost}@{}}%
\column{8}{@{}>{\hspre}l<{\hspost}@{}}%
\column{E}{@{}>{\hspre}l<{\hspost}@{}}%
\>[B]{}\Varid{prod}_{1}{}\<[8]%
\>[8]{}\mathrel{=}\Varid{prod}\mathbin{\circ}\Conid{Wrap}{}\<[E]%
\\
\>[B]{}\Varid{prod}_{2}{}\<[8]%
\>[8]{}\mathrel{=}\Varid{prod}\mathbin{\circ}\Conid{Return}{}\<[E]%
\ColumnHook
\end{hscode}\resethooks
The definition of \ensuremath{\Varid{prod}_{2}} is forced:
\begin{hscode}\SaveRestoreHook
\column{B}{@{}>{\hspre}c<{\hspost}@{}}%
\column{BE}{@{}l@{}}%
\column{3}{@{}>{\hspre}l<{\hspost}@{}}%
\column{5}{@{}>{\hspre}l<{\hspost}@{}}%
\column{E}{@{}>{\hspre}l<{\hspost}@{}}%
\>[3]{}\Varid{prod}_{2}{}\<[E]%
\\
\>[B]{}\mathrel{=}{}\<[BE]%
\>[5]{}\mbox{\commentbegin  definition of \ensuremath{\Varid{prod}_{2}}  \commentend}{}\<[E]%
\\
\>[B]{}\hsindent{3}{}\<[3]%
\>[3]{}\Varid{prod}\mathbin{\circ}\Conid{Return}{}\<[E]%
\\
\>[B]{}\mathrel{=}{}\<[BE]%
\>[5]{}\mbox{\commentbegin  \ensuremath{\Conid{F}} as a premonad: \ensuremath{\Varid{return}_{\Conid{F}}\mathrel{=}\Conid{Return}}  \commentend}{}\<[E]%
\\
\>[B]{}\hsindent{3}{}\<[3]%
\>[3]{}\Varid{prod}\mathbin{\circ}\Varid{return}_{\Conid{F}}{}\<[E]%
\\
\>[B]{}\mathrel{=}{}\<[BE]%
\>[5]{}\mbox{\commentbegin  property (1)  \commentend}{}\<[E]%
\\
\>[B]{}\hsindent{3}{}\<[3]%
\>[3]{}\Varid{id}{}\<[E]%
\ColumnHook
\end{hscode}\resethooks
Now consider property (2). Expanding the left-hand side, we have
\begin{hscode}\SaveRestoreHook
\column{B}{@{}>{\hspre}c<{\hspost}@{}}%
\column{BE}{@{}l@{}}%
\column{3}{@{}>{\hspre}l<{\hspost}@{}}%
\column{5}{@{}>{\hspre}l<{\hspost}@{}}%
\column{E}{@{}>{\hspre}l<{\hspost}@{}}%
\>[3]{}\Varid{prod}\mathbin{\circ}\Conid{F}\;\Varid{return}{}\<[E]%
\\
\>[B]{}\mathrel{=}{}\<[BE]%
\>[5]{}\mbox{\commentbegin  datatype isomorphism  \commentend}{}\<[E]%
\\
\>[B]{}\hsindent{3}{}\<[3]%
\>[3]{}\Varid{prod}\mathbin{\circ}\Conid{F}\;\Varid{return}\mathbin{\circ}(\Conid{Wrap}\mathbin{\triangledown}\Conid{Return})\mathbin{\circ}\Varid{out}_{\Conid{F}}{}\<[E]%
\\
\>[B]{}\mathrel{=}{}\<[BE]%
\>[5]{}\mbox{\commentbegin  naturality of \ensuremath{\Conid{Wrap}\mathbin{\triangledown}\Conid{Return}}  \commentend}{}\<[E]%
\\
\>[B]{}\hsindent{3}{}\<[3]%
\>[3]{}\Varid{prod}\mathbin{\circ}(\Conid{Wrap}\mathbin{\triangledown}\Conid{Return})\mathbin{\circ}(\Conid{M}\;\Conid{F}\;\Varid{return}\mathbin{+}\Varid{return})\mathbin{\circ}\Varid{out}_{\Conid{F}}{}\<[E]%
\\
\>[B]{}\mathrel{=}{}\<[BE]%
\>[5]{}\mbox{\commentbegin  definitions of \ensuremath{\Varid{prod}_{1}}, \ensuremath{\Varid{prod}_{2}}  \commentend}{}\<[E]%
\\
\>[B]{}\hsindent{3}{}\<[3]%
\>[3]{}(\Varid{prod}_{1}\mathbin{\triangledown}\Varid{prod}_{2})\mathbin{\circ}(\Conid{M}\;\Conid{F}\;\Varid{return}\mathbin{+}\Varid{return})\mathbin{\circ}\Varid{out}_{\Conid{F}}{}\<[E]%
\\
\>[B]{}\mathrel{=}{}\<[BE]%
\>[5]{}\mbox{\commentbegin  coproducts  \commentend}{}\<[E]%
\\
\>[B]{}\hsindent{3}{}\<[3]%
\>[3]{}((\Varid{prod}_{1}\mathbin{\circ}\Conid{M}\;\Conid{F}\;\Varid{return})\mathbin{\triangledown}(\Varid{prod}_{2}\mathbin{\circ}\Varid{return}))\mathbin{\circ}\Varid{out}_{\Conid{F}}{}\<[E]%
\ColumnHook
\end{hscode}\resethooks
and for the right-hand side we have
\begin{hscode}\SaveRestoreHook
\column{B}{@{}>{\hspre}c<{\hspost}@{}}%
\column{BE}{@{}l@{}}%
\column{3}{@{}>{\hspre}l<{\hspost}@{}}%
\column{5}{@{}>{\hspre}l<{\hspost}@{}}%
\column{E}{@{}>{\hspre}l<{\hspost}@{}}%
\>[3]{}\Varid{return}_{\Conid{M}}{}\<[E]%
\\
\>[B]{}\mathrel{=}{}\<[BE]%
\>[5]{}\mbox{\commentbegin  datatype isomorphism  \commentend}{}\<[E]%
\\
\>[B]{}\hsindent{3}{}\<[3]%
\>[3]{}\Varid{return}_{\Conid{M}}\mathbin{\circ}(\Conid{Wrap}\mathbin{\triangledown}\Conid{Return})\mathbin{\circ}\Varid{out}_{\Conid{F}}{}\<[E]%
\\
\>[B]{}\mathrel{=}{}\<[BE]%
\>[5]{}\mbox{\commentbegin  coproducts  \commentend}{}\<[E]%
\\
\>[B]{}\hsindent{3}{}\<[3]%
\>[3]{}((\Varid{return}_{\Conid{M}}\mathbin{\circ}\Conid{Wrap})\mathbin{\triangledown}(\Varid{return}_{\Conid{M}}\mathbin{\circ}\Conid{Return}))\mathbin{\circ}\Varid{out}_{\Conid{F}}{}\<[E]%
\ColumnHook
\end{hscode}\resethooks
These two expressions must be equal, which implies that the equalities
\begin{hscode}\SaveRestoreHook
\column{B}{@{}>{\hspre}l<{\hspost}@{}}%
\column{21}{@{}>{\hspre}l<{\hspost}@{}}%
\column{E}{@{}>{\hspre}l<{\hspost}@{}}%
\>[B]{}\Varid{prod}_{1}\mathbin{\circ}\Conid{M}\;\Conid{F}\;\Varid{return}{}\<[21]%
\>[21]{}\mathrel{=}\Varid{return}_{\Conid{M}}\mathbin{\circ}\Conid{Wrap}{}\<[E]%
\\
\>[B]{}\Varid{prod}_{2}\mathbin{\circ}\Varid{return}{}\<[21]%
\>[21]{}\mathrel{=}\Varid{return}_{\Conid{M}}\mathbin{\circ}\Conid{Return}{}\<[E]%
\ColumnHook
\end{hscode}\resethooks
must hold. The second equality follows from \ensuremath{\Varid{prod}_{2}\mathrel{=}\Varid{id}} and the definition of \ensuremath{\Varid{return}}; we'll use the first one to derive a definition for \ensuremath{\Varid{prod}_{1}}. We have:
\begin{hscode}\SaveRestoreHook
\column{B}{@{}>{\hspre}c<{\hspost}@{}}%
\column{BE}{@{}l@{}}%
\column{3}{@{}>{\hspre}l<{\hspost}@{}}%
\column{5}{@{}>{\hspre}l<{\hspost}@{}}%
\column{E}{@{}>{\hspre}l<{\hspost}@{}}%
\>[3]{}\Varid{prod}_{1}\mathbin{\circ}\Conid{M}\;\Conid{F}\;\Varid{return}{}\<[E]%
\\
\>[B]{}\mathrel{=}{}\<[BE]%
\>[5]{}\mbox{\commentbegin  (A) suppose that \ensuremath{\Varid{prod}_{1}\mathrel{=}\Varid{prod}_{1}'\mathbin{\circ}\Conid{M}\;\Varid{prod}}  \commentend}{}\<[E]%
\\
\>[B]{}\hsindent{3}{}\<[3]%
\>[3]{}\Varid{prod}_{1}'\mathbin{\circ}\Conid{M}\;\Varid{prod}\mathbin{\circ}\Conid{M}\;\Conid{F}\;\Varid{return}{}\<[E]%
\\
\>[B]{}\mathrel{=}{}\<[BE]%
\>[5]{}\mbox{\commentbegin  functors  \commentend}{}\<[E]%
\\
\>[B]{}\hsindent{3}{}\<[3]%
\>[3]{}\Varid{prod}_{1}'\mathbin{\circ}\Conid{M}\;(\Varid{prod}\mathbin{\circ}\Conid{F}\;\Varid{return}){}\<[E]%
\\
\>[B]{}\mathrel{=}{}\<[BE]%
\>[5]{}\mbox{\commentbegin  (B) induction  \commentend}{}\<[E]%
\\
\>[B]{}\hsindent{3}{}\<[3]%
\>[3]{}\Varid{prod}_{1}'\mathbin{\circ}\Conid{M}\;\Varid{return}_{\Conid{M}}{}\<[E]%
\\
\>[B]{}\mathrel{=}{}\<[BE]%
\>[5]{}\mbox{\commentbegin  (C) suppose that \ensuremath{\Varid{prod}_{1}'\mathrel{=}\Varid{return}_{\Conid{M}}\mathbin{\circ}\Conid{Wrap}\mathbin{\circ}\Varid{join}_{\Conid{M}}}  \commentend}{}\<[E]%
\\
\>[B]{}\hsindent{3}{}\<[3]%
\>[3]{}\Varid{return}_{\Conid{M}}\mathbin{\circ}\Conid{Wrap}\mathbin{\circ}\Varid{join}_{\Conid{M}}\mathbin{\circ}\Conid{M}\;\Varid{return}_{\Conid{M}}{}\<[E]%
\\
\>[B]{}\mathrel{=}{}\<[BE]%
\>[5]{}\mbox{\commentbegin  \ensuremath{\Conid{M}} is a monad  \commentend}{}\<[E]%
\\
\>[B]{}\hsindent{3}{}\<[3]%
\>[3]{}\Varid{return}_{\Conid{M}}\mathbin{\circ}\Conid{Wrap}{}\<[E]%
\ColumnHook
\end{hscode}\resethooks
and so property (2) strongly suggests letting
\begin{hscode}\SaveRestoreHook
\column{B}{@{}>{\hspre}l<{\hspost}@{}}%
\column{E}{@{}>{\hspre}l<{\hspost}@{}}%
\>[B]{}\Varid{prod}_{1}\mathrel{=}\Varid{return}_{\Conid{M}}\mathbin{\circ}\Conid{Wrap}\mathbin{\circ}\Varid{join}_{\Conid{M}}\mathbin{\circ}\Conid{M}\;\Varid{prod}{}\<[E]%
\ColumnHook
\end{hscode}\resethooks
The three steps (A), (B), (C) need a little justification. For (A),
we're heading towards a use of induction, so we require an occurrence
of \ensuremath{\Conid{M}\;\Varid{prod}} at the end of \ensuremath{\Varid{prod}_{1}}. For (B), induction seems plausible,
because this calculation takes places under a \ensuremath{\Conid{Wrap}} constructor. 
For (C), the final \ensuremath{\Conid{M}\;\Varid{return}_{\Conid{M}}} is cancellable via
\ensuremath{\Varid{join}_{\Conid{M}}}, so we're done---we let be \ensuremath{\Varid{prod}_{1}'} be the final expression
we want composed with this cancellation.

We don't need any stronger justification for induction than mere
plausibility, because we are using this calculation only to suggest a
possible definition of \ensuremath{\Varid{prod}_{1}} that we should then check more
rigorously.  This still leaves us also with having to check property (3),
again by induction. Both of these proofs are presented in \ref{sec:self-monad-proof}.

The final instance declaration for \ensuremath{\mathit{Nest}} is then as follows:

\begin{hscode}\SaveRestoreHook
\column{B}{@{}>{\hspre}l<{\hspost}@{}}%
\column{3}{@{}>{\hspre}l<{\hspost}@{}}%
\column{5}{@{}>{\hspre}l<{\hspost}@{}}%
\column{7}{@{}>{\hspre}l<{\hspost}@{}}%
\column{11}{@{}>{\hspre}l<{\hspost}@{}}%
\column{31}{@{}>{\hspre}l<{\hspost}@{}}%
\column{E}{@{}>{\hspre}l<{\hspost}@{}}%
\>[B]{}\mathbf{instance}\;(\Conid{Functor}\;\Varid{m},\ \Conid{Monad}\;\Varid{m})\Rightarrow \Conid{Monad}\;(\mathit{Nest}\;\Varid{m})\;\mathbf{where}{}\<[E]%
\\
\>[B]{}\hsindent{3}{}\<[3]%
\>[3]{}\Varid{return}{}\<[11]%
\>[11]{}\mathrel{=}\mathit{Nest}\mathbin{\circ}\Varid{return}_{\Conid{M}}\mathbin{\circ}\Conid{Return}{}\<[E]%
\\[\blanklineskip]%
\>[B]{}\hsindent{3}{}\<[3]%
\>[3]{}\Varid{join}\mathrel{=}\mathit{Nest}\mathbin{\circ}\Varid{join}_{\Conid{M}}\mathbin{\circ}\Varid{fmap}_{\Conid{M}}\;\Varid{prod}\mathbin{\circ}\Varid{unNest}{}\<[E]%
\\
\>[3]{}\hsindent{2}{}\<[5]%
\>[5]{}\mathbf{where}{}\<[E]%
\\
\>[5]{}\hsindent{2}{}\<[7]%
\>[7]{}\Varid{prod}\;(\Conid{Return}\;(\mathit{Nest}\;\Varid{m})){}\<[31]%
\>[31]{}\mathrel{=}\Varid{m}{}\<[E]%
\\
\>[5]{}\hsindent{2}{}\<[7]%
\>[7]{}\Varid{prod}\;(\Conid{Wrap}\;\Varid{m}){}\<[31]%
\>[31]{}\mathrel{=}(\Varid{return}_{\Conid{M}}\mathbin{\circ}\Conid{Wrap}\mathbin{\circ}\Varid{join}_{\Conid{M}}\mathbin{\circ}\Varid{fmap}_{\Conid{M}}\;\Varid{prod})\;\Varid{m}{}\<[E]%
\ColumnHook
\end{hscode}\resethooks

\subsection{Tracing with \ensuremath{\mathit{Nest}}} \label{sec:nest-tracer}

For any monad \ensuremath{\Varid{m}}, the monad \ensuremath{\mathit{Nest}\;\Varid{m}} is a tracer. The \ensuremath{\Varid{lift}} function maps \ensuremath{\Conid{Return}} on values. It does not create any \ensuremath{\Conid{Wrap}} constructor, so lifted computations are single-level and are always joined when \ensuremath{\Varid{join}} for \ensuremath{\mathit{Nest}} is performed. The \ensuremath{\Varid{drop}} function traverses the free structure, and collapses levels which were previously separated by \ensuremath{\Conid{Wrap}} constructors. The \ensuremath{\Varid{mark}} gadget explicitly creates a new level by the \ensuremath{\Conid{Wrap}} constructor mapped on \ensuremath{\Conid{Return}}s. All future computations will be confined to the wrapped \ensuremath{\Conid{Return}}s, and so the previous structure is preserved.
\begin{hscode}\SaveRestoreHook
\column{B}{@{}>{\hspre}l<{\hspost}@{}}%
\column{3}{@{}>{\hspre}l<{\hspost}@{}}%
\column{5}{@{}>{\hspre}l<{\hspost}@{}}%
\column{7}{@{}>{\hspre}l<{\hspost}@{}}%
\column{14}{@{}>{\hspre}l<{\hspost}@{}}%
\column{26}{@{}>{\hspre}l<{\hspost}@{}}%
\column{E}{@{}>{\hspre}l<{\hspost}@{}}%
\>[B]{}\mathbf{instance}\;(\Conid{Functor}\;\Varid{m},\ \Conid{Monad}\;\Varid{m})\Rightarrow \Conid{MonadTrans}\;\Varid{m}\;(\mathit{Nest}\;\Varid{m})\;\mathbf{where}{}\<[E]%
\\
\>[B]{}\hsindent{3}{}\<[3]%
\>[3]{}\Varid{lift}{}\<[14]%
\>[14]{}\mathrel{=}\mathit{Nest}\mathbin{\circ}\Varid{fmap}\;\Conid{Return}{}\<[E]%
\\[\blanklineskip]%
\>[B]{}\hsindent{3}{}\<[3]%
\>[3]{}\Varid{drop}\;\Varid{v}{}\<[14]%
\>[14]{}\mathrel{=}\mathit{unNest}\;\Varid{v}\bind \Varid{revert}{}\<[E]%
\\
\>[3]{}\hsindent{2}{}\<[5]%
\>[5]{}\mathbf{where}{}\<[E]%
\\
\>[5]{}\hsindent{2}{}\<[7]%
\>[7]{}\Varid{revert}\;(\Conid{Return}\;\Varid{a}){}\<[26]%
\>[26]{}\mathrel{=}\Varid{return}\;\Varid{a}{}\<[E]%
\\
\>[5]{}\hsindent{2}{}\<[7]%
\>[7]{}\Varid{revert}\;(\Conid{Wrap}\;\Varid{m}){}\<[26]%
\>[26]{}\mathrel{=}\Varid{m}\bind \Varid{revert}{}\<[E]%
\\[\blanklineskip]%
\>[B]{}\mathbf{instance}\;(\Conid{Functor}\;\Varid{m},\ \Conid{Monad}\;\Varid{m})\Rightarrow \Conid{MonadTrace}\;(\mathit{Nest}\;\Varid{m})\;\mathbf{where}{}\<[E]%
\\
\>[B]{}\hsindent{3}{}\<[3]%
\>[3]{}\Varid{mark}{}\<[14]%
\>[14]{}\mathrel{=}(\mathit{Nest}\mathbin{\circ}\Varid{return}\mathbin{\circ}\Conid{Wrap}\mathbin{\circ}\Varid{return}\mathbin{\circ}\Conid{Return})\;(){}\<[E]%
\ColumnHook
\end{hscode}\resethooks
The proof that these definitions satisfy the laws of \ensuremath{\Conid{MonadTrace}} from Section~\ref{sec:tracer} is given in \ref{sec:appendix}.

As an example of \ensuremath{\Varid{mark}}, consider the following conversation with the Haskell shell. The call of the \ensuremath{\Varid{mark}} operation in the second query maps \ensuremath{\Conid{Wrap}} on every element. This way, every future computation is confined to the inside of those two \ensuremath{\Conid{Wrap}} constructors. Whatever monadic computation is bound to the result, on the top-level there will always be a two-element list, because it is a node with only \ensuremath{\Conid{Wrap}} constructors as children.
\begin{hscode}\SaveRestoreHook
\column{B}{@{}>{\hspre}l<{\hspost}@{}}%
\column{E}{@{}>{\hspre}l<{\hspost}@{}}%
\>[B]{}\mathtt{\triangleright}\;\Varid{lift}\;[\mskip1.5mu \mathrm{1},\mathrm{2}\mskip1.5mu]{}\<[E]%
\\
\>[B]{}\mathit{Nest}\;[\mskip1.5mu \Conid{Return}\;\mathrm{1},\Conid{Return}\;\mathrm{2}\mskip1.5mu]{}\<[E]%
\\[\blanklineskip]%
\>[B]{}\mathtt{\triangleright}\;\mathbf{do}\;\{\mskip1.5mu \Varid{x}\leftarrow \Varid{lift}\;[\mskip1.5mu \mathrm{1},\mathrm{2}\mskip1.5mu];\ \Varid{lift}\;[\mskip1.5mu \mathrm{0}\mathinner{\ldotp\ldotp}\Varid{x}\mskip1.5mu]\mskip1.5mu\}{}\<[E]%
\\
\>[B]{}\mathit{Nest}\;[\mskip1.5mu \Conid{Return}\;\mathrm{0},\Conid{Return}\;\mathrm{1},\Conid{Return}\;\mathrm{0},\Conid{Return}\;\mathrm{1},\Conid{Return}\;\mathrm{2}\mskip1.5mu]{}\<[E]%
\\[\blanklineskip]%
\>[B]{}\mathtt{\triangleright}\;\mathbf{do}\;\{\mskip1.5mu \Varid{lift}\;[\mskip1.5mu \mathrm{1},\mathrm{2}\mskip1.5mu];\ \Varid{mark}\mskip1.5mu\}{}\<[E]%
\\
\>[B]{}\mathit{Nest}\;[\mskip1.5mu \Conid{Wrap}\;[\mskip1.5mu \Conid{Return}\;()\mskip1.5mu],\Conid{Wrap}\;[\mskip1.5mu \Conid{Return}\;()\mskip1.5mu]\mskip1.5mu]{}\<[E]%
\\[\blanklineskip]%
\>[B]{}\mathtt{\triangleright}\;\mathbf{do}\;\{\mskip1.5mu \Varid{x}\leftarrow \Varid{lift}\;[\mskip1.5mu \mathrm{1},\mathrm{2}\mskip1.5mu];\ \Varid{mark};\ \Varid{lift}\;[\mskip1.5mu \mathrm{0}\mathinner{\ldotp\ldotp}\Varid{x}\mskip1.5mu]\mskip1.5mu\}{}\<[E]%
\\
\>[B]{}\mathit{Nest}\;[\mskip1.5mu \Conid{Wrap}\;[\mskip1.5mu \Conid{Return}\;\mathrm{0},\Conid{Return}\;\mathrm{1}\mskip1.5mu],\Conid{Wrap}\;[\mskip1.5mu \Conid{Return}\;\mathrm{0},\Conid{Return}\;\mathrm{1},\Conid{Return}\;\mathrm{2}\mskip1.5mu]\mskip1.5mu]{}\<[E]%
\ColumnHook
\end{hscode}\resethooks

\section{Interpreting traces} \label{sec:traces}

The tracing construction has multiple applications. As in the motivating example from the introduction, we can track the course of computation, in order to identify any point of failure. But, stepping back from this particular example, we observe that a trace is simply a data structure, and the computation in the \ensuremath{\mathit{Nest}} monad is performed only if the data structure is forced. This gives us control over the process of execution, which can be useful in the context of non-terminating computations (although the inductive proofs will need strengthening in that context). We can interpret the free parts of a \ensuremath{\mathit{Nest}} value in any way we want, and thus mix in some new effects, not previously available with the original monad. This control over a computation from the outside strongly resembles the paradigm of aspect-oriented programming. 

\subsection{The partiality monad}\label{sec:partiality}

Capretta introduced the partiality monad~\cite{capretta:2005} to capture non-termination as an effect; this technique has applications in type theory, to model non-guarded recursion. The original formulation is as follows.
\begin{hscode}\SaveRestoreHook
\column{B}{@{}>{\hspre}l<{\hspost}@{}}%
\column{E}{@{}>{\hspre}l<{\hspost}@{}}%
\>[B]{}\mathbf{data}\;\Conid{Partial}\;\Varid{a}\mathrel{=}\Conid{Later}\;(\Conid{Partial}\;\Varid{a})\mid \Conid{Now}\;\Varid{a}{}\<[E]%
\ColumnHook
\end{hscode}\resethooks
A structure of this type is a value wrapped in a (possibly infinite) number of \ensuremath{\Conid{Later}} constructors. It represents a computation sliced into layers. We can explicitly force any number of layers.
The \ensuremath{\mathit{Nest}} monad can be seen as a monad transformer which allows computations structured by any monad to be sliced. The most basic case, \ensuremath{\mathit{Nest}\;\Conid{Identity}}, is indeed isomorphic to \ensuremath{\Conid{Partial}}.

In most programming languages, including Haskell, the \ensuremath{\mathrel{\vee}} operator is asymmetric---non-strict in its second argument (\ensuremath{\Conid{True}\mathrel{\vee}\bot \equiv \Conid{True}}) but strict in its first (\ensuremath{\bot \mathrel{\vee}\Conid{True}\equiv \bot }). In pure Haskell, it is impossible to define \emph{parallel-or}---that is, a disjunction \ensuremath{\curlyvee} with the property that \ensuremath{\Conid{True}\curlyvee\bot \equiv \Conid{True}\equiv \bot \curlyvee\Conid{True}}.

To implement such a disjunction in the real world, we need some kind of parallelism, so that both arguments are evaluated simultaneously; when either one terminates with \ensuremath{\Conid{True}} or both terminate with \ensuremath{\Conid{False}}, the computation of the disjunction is complete. We can purely approximate this behaviour---at least for the case where undefined arguments arise from non-termination, rather than from any other reason. We do this by explicitly cutting the infinite computation into finite pieces, using the \ensuremath{\mathit{Nest}} tracer.

Consider the following function, which is an implementation of the so-called Collatz problem. It is suspected that for all $n>0$ it returns \ensuremath{\Conid{True}}, but no proof for this claim is known.
\begin{hscode}\SaveRestoreHook
\column{B}{@{}>{\hspre}l<{\hspost}@{}}%
\column{12}{@{}>{\hspre}l<{\hspost}@{}}%
\column{25}{@{}>{\hspre}l<{\hspost}@{}}%
\column{E}{@{}>{\hspre}l<{\hspost}@{}}%
\>[B]{}\Varid{collatz}\mathbin{::}\Conid{Integer}\to \Conid{Bool}{}\<[E]%
\\
\>[B]{}\Varid{collatz}\;\mathrm{1}{}\<[25]%
\>[25]{}\mathrel{=}\Conid{True}{}\<[E]%
\\
\>[B]{}\Varid{collatz}\;\Varid{n}{}\<[12]%
\>[12]{}\mid \Varid{odd}\;\Varid{n}{}\<[25]%
\>[25]{}\mathrel{=}\Varid{collatz}\;(\mathrm{3}\times\Varid{n}\mathbin{+}\mathrm{1}){}\<[E]%
\\
\>[12]{}\mid \Varid{otherwise}{}\<[25]%
\>[25]{}\mathrel{=}\Varid{collatz}\;(\Varid{n}\div\mathrm{2}){}\<[E]%
\ColumnHook
\end{hscode}\resethooks

\noindent
If we want to check whether at least one of two Collatz sequences ends, it is not the best idea to use the regular Haskell disjunction, since if the first one diverges, the whole function diverges too.

\begin{hscode}\SaveRestoreHook
\column{B}{@{}>{\hspre}l<{\hspost}@{}}%
\column{E}{@{}>{\hspre}l<{\hspost}@{}}%
\>[B]{}\Varid{oneOf}\mathbin{::}\Conid{Integer}\to \Conid{Integer}\to \Conid{Bool}{}\<[E]%
\\
\>[B]{}\Varid{oneOf}\;\Varid{a}\;\Varid{b}\mathrel{=}\Varid{collatz}\;\Varid{a}\mathrel{\vee}\Varid{collatz}\;\Varid{b}{}\<[E]%
\ColumnHook
\end{hscode}\resethooks

A much safer solution is to chop the evaluation of the Collatz sequence into pieces. We use the \ensuremath{\mathit{Nest}} tracer for the \ensuremath{\Conid{Identity}} monad. This way, we make the evaluation incremental, which enables us to execute it in parallel. The scheduler is very simple and hidden in the definition of \ensuremath{\curlyvee}. It executes a piece from each argument in turn.

\begin{hscode}\SaveRestoreHook
\column{B}{@{}>{\hspre}l<{\hspost}@{}}%
\column{13}{@{}>{\hspre}l<{\hspost}@{}}%
\column{26}{@{}>{\hspre}l<{\hspost}@{}}%
\column{31}{@{}>{\hspre}l<{\hspost}@{}}%
\column{38}{@{}>{\hspre}l<{\hspost}@{}}%
\column{39}{@{}>{\hspre}l<{\hspost}@{}}%
\column{69}{@{}>{\hspre}l<{\hspost}@{}}%
\column{E}{@{}>{\hspre}l<{\hspost}@{}}%
\>[B]{}\Varid{collatzN}\mathbin{::}\Conid{Integer}\to \mathit{Nest}\;\Conid{Identity}\;\Conid{Bool}{}\<[E]%
\\
\>[B]{}\Varid{collatzN}\;\mathrm{1}{}\<[26]%
\>[26]{}\mathrel{=}\Varid{return}\;{}\<[38]%
\>[38]{}\Conid{True}{}\<[E]%
\\
\>[B]{}\Varid{collatzN}\;\Varid{n}{}\<[13]%
\>[13]{}\mid \Varid{odd}\;\Varid{n}{}\<[26]%
\>[26]{}\mathrel{=}\Varid{mark}\sequ {}\<[38]%
\>[38]{}\Varid{collatzN}\;(\mathrm{3}\times\Varid{n}\mathbin{+}\mathrm{1}){}\<[E]%
\\
\>[13]{}\mid \Varid{otherwise}{}\<[26]%
\>[26]{}\mathrel{=}\Varid{mark}\sequ {}\<[38]%
\>[38]{}\Varid{collatzN}\;(\Varid{n}\div\mathrm{2}){}\<[E]%
\\[\blanklineskip]%
\>[B]{}(\curlyvee)\mathbin{::}\mathit{Nest}\;\Conid{Identity}\;\Conid{Bool}\to \mathit{Nest}\;\Conid{Identity}\;\Conid{Bool}\to \Conid{Bool}{}\<[E]%
\\
\>[B]{}\mathit{Nest}\;(\Conid{Identity}\;(\Conid{Leaf}\;\Conid{False})){}\<[31]%
\>[31]{}\curlyvee{}\<[39]%
\>[39]{}\mathit{Nest}\;(\Conid{Identity}\;(\Conid{Leaf}\;\Conid{False})){}\<[69]%
\>[69]{}\mathrel{=}\Conid{False}{}\<[E]%
\\
\>[B]{}\mathit{Nest}\;(\Conid{Identity}\;(\Conid{Leaf}\;\Conid{True})){}\<[31]%
\>[31]{}\curlyvee{}\<[39]%
\>[39]{}\anonymous {}\<[69]%
\>[69]{}\mathrel{=}\Conid{True}{}\<[E]%
\\
\>[B]{}\anonymous {}\<[31]%
\>[31]{}\curlyvee{}\<[39]%
\>[39]{}\mathit{Nest}\;(\Conid{Identity}\;(\Conid{Leaf}\;\Conid{True})){}\<[69]%
\>[69]{}\mathrel{=}\Conid{True}{}\<[E]%
\\
\>[B]{}\mathit{Nest}\;(\Conid{Identity}\;(\Conid{Node}\;\Varid{m})){}\<[31]%
\>[31]{}\curlyvee{}\<[39]%
\>[39]{}\Varid{x}{}\<[69]%
\>[69]{}\mathrel{=}\Varid{x}\curlyvee\mathit{Nest}\;\Varid{m}{}\<[E]%
\ColumnHook
\end{hscode}\resethooks

\noindent
We can test the \ensuremath{\curlyvee} operator as follows. Note that \ensuremath{\Varid{collatz}} diverges if applied to \ensuremath{\mathrm{0}}.

\begin{hscode}\SaveRestoreHook
\column{B}{@{}>{\hspre}l<{\hspost}@{}}%
\column{E}{@{}>{\hspre}l<{\hspost}@{}}%
\>[B]{}\mathtt{\triangleright}\;\Varid{collatzN}\;\mathrm{120}\curlyvee\Varid{collatzN}\;\mathrm{130}{}\<[E]%
\\
\>[B]{}\Conid{True}{}\<[E]%
\\[\blanklineskip]%
\>[B]{}\mathtt{\triangleright}\;\Varid{collatzN}\;\mathrm{0}\curlyvee\Varid{collatzN}\;\mathrm{130}{}\<[E]%
\\
\>[B]{}\Conid{True}{}\<[E]%
\\[\blanklineskip]%
\>[B]{}\mathtt{\triangleright}\;\Varid{collatzN}\;\mathrm{130}\curlyvee\Varid{collatzN}\;\mathrm{0}{}\<[E]%
\\
\>[B]{}\Conid{True}{}\<[E]%
\ColumnHook
\end{hscode}\resethooks

\noindent
This model can easily be extended to different kinds of such thread races. For example, it is possible to simulate McCarthy's ambiguous choice operator \ensuremath{\Varid{amb}\mathbin{::}\Varid{a}\to \Varid{b}\to \Conid{Either}\;\Varid{a}\;\Varid{b}} \cite{McCarthy63:Basis}, which has the property that for \ensuremath{\Varid{a}_{\mathrm{0}}\mathbin{::}\Varid{a}} and \ensuremath{\Varid{b}_{\mathrm{0}}\mathbin{::}\Varid{b}}, \ensuremath{\Varid{amb}\;\Varid{a}_{\mathrm{0}}\;\bot \mathrel{=}\Conid{Left}\;\Varid{a}_{\mathrm{0}}}, and \ensuremath{\Varid{amb}\;\bot \;\Varid{b}_{\mathrm{0}}\mathrel{=}\Conid{Right}\;\Varid{b}_{\mathrm{0}}} (again, assuming that the undefined values arise from non-termination).

\subsection{Approximating computations}

For some monads, the initial segment of a \ensuremath{\mathit{Nest}} value may be seen as an ``approximation'' to the computation. Consider the monad of finitely supported probability distributions. Its most common representation is a list of probabilities paired with values.
\begin{hscode}\SaveRestoreHook
\column{B}{@{}>{\hspre}l<{\hspost}@{}}%
\column{3}{@{}>{\hspre}l<{\hspost}@{}}%
\column{54}{@{}>{\hspre}l<{\hspost}@{}}%
\column{E}{@{}>{\hspre}l<{\hspost}@{}}%
\>[B]{}\mathbf{newtype}\;\Conid{Distr}\;\Varid{a}\mathrel{=}\Conid{Distr}\{\mskip1.5mu \Varid{runDistr}\mathbin{::}[\mskip1.5mu (\Conid{Double},\Varid{a})\mskip1.5mu]\mskip1.5mu\}{}\<[E]%
\\[\blanklineskip]%
\>[B]{}\mathbf{instance}\;\Conid{Functor}\;\Conid{Distr}\;\mathbf{where}{}\<[E]%
\\
\>[B]{}\hsindent{3}{}\<[3]%
\>[3]{}\Varid{fmap}\;\Varid{f}\;(\Conid{Distr}\;\Varid{xs})\mathrel{=}\Conid{Distr}\;(\Varid{fmap}\;(\lambda (\Varid{p},\Varid{a})\to (\Varid{p},\ {}\<[54]%
\>[54]{}\Varid{f}\;\Varid{a}))\;\Varid{xs}){}\<[E]%
\\[\blanklineskip]%
\>[B]{}\mathbf{instance}\;\Conid{Monad}\;\Conid{Distr}\;\mathbf{where}{}\<[E]%
\\
\>[B]{}\hsindent{3}{}\<[3]%
\>[3]{}\Varid{return}\;\Varid{a}\mathrel{=}\Conid{Distr}\;[\mskip1.5mu (\mathrm{1},\Varid{a})\mskip1.5mu]{}\<[E]%
\\
\>[B]{}\hsindent{3}{}\<[3]%
\>[3]{}\Varid{join}\;(\Conid{Distr}\;\Varid{xs})\mathrel{=}\Conid{Distr}\;[\mskip1.5mu (\Varid{p}\times\Varid{q},\Varid{a})\mid (\Varid{p},\Conid{Distr}\;\Varid{ys})\leftarrow \Varid{xs},\ (\Varid{q},\Varid{a})\leftarrow \Varid{ys}\mskip1.5mu]{}\<[E]%
\ColumnHook
\end{hscode}\resethooks
However, this representation has a flaw. Consider the following problem: given the uniform distribution of $\{0,1\}$ (a~fair coin), select uniformly an element from $\{0,1,2\}$. A solution is to flip the coin twice to get the uniform distribution of $\{0,1,2,3\}$; if you draw $0$, $1$, or $2$, this is your answer, and if you draw $3$, flip twice more. In Haskell:
\begin{hscode}\SaveRestoreHook
\column{B}{@{}>{\hspre}l<{\hspost}@{}}%
\column{8}{@{}>{\hspre}l<{\hspost}@{}}%
\column{14}{@{}>{\hspre}l<{\hspost}@{}}%
\column{16}{@{}>{\hspre}l<{\hspost}@{}}%
\column{E}{@{}>{\hspre}l<{\hspost}@{}}%
\>[B]{}\Varid{coin}{}\<[8]%
\>[8]{}\mathbin{::}\Conid{Distr}\;\Conid{Int}{}\<[E]%
\\
\>[B]{}\Varid{coin}{}\<[8]%
\>[8]{}\mathrel{=}\Conid{Distr}\;[\mskip1.5mu (\mathrm{0.5},\mathrm{0}),(\mathrm{0.5},\mathrm{1})\mskip1.5mu]{}\<[E]%
\\[\blanklineskip]%
\>[B]{}\Varid{third}{}\<[8]%
\>[8]{}\mathbin{::}\Conid{Distr}\;\Conid{Int}{}\<[E]%
\\
\>[B]{}\Varid{third}{}\<[8]%
\>[8]{}\mathrel{=}\mathbf{do}\;{}\<[14]%
\>[14]{}\Varid{x}\leftarrow \Varid{coin}{}\<[E]%
\\
\>[14]{}\Varid{y}\leftarrow \Varid{coin}{}\<[E]%
\\
\>[14]{}\mathbf{case}\;(\Varid{x},\Varid{y})\;\mathbf{of}{}\<[E]%
\\
\>[14]{}\hsindent{2}{}\<[16]%
\>[16]{}(\mathrm{1},\mathrm{1})\to \Varid{return}\;\mathrm{0}{}\<[E]%
\\
\>[14]{}\hsindent{2}{}\<[16]%
\>[16]{}(\mathrm{1},\mathrm{0})\to \Varid{return}\;\mathrm{1}{}\<[E]%
\\
\>[14]{}\hsindent{2}{}\<[16]%
\>[16]{}(\mathrm{0},\mathrm{1})\to \Varid{return}\;\mathrm{2}{}\<[E]%
\\
\>[14]{}\hsindent{2}{}\<[16]%
\>[16]{}(\mathrm{0},\mathrm{0})\to \Varid{third}{}\<[E]%
\ColumnHook
\end{hscode}\resethooks
(This is a simplification of Knuth and Yao's technique to simulate a fair die using three fair coin tosses \cite{KY76}.)
Though the solution is mathematically reasonable, the Haskell implementation is useless, because the \ensuremath{\Conid{List}} monad, and so also the \ensuremath{\Conid{Distr}} monad, gathers the results in a depth-first fashion. Though this may not be obvious at first sight, the recursive call in \ensuremath{\Varid{third}} is in the head of the list. Therefore, \ensuremath{\Varid{third}} actually diverges without producing any usable results: \ensuremath{\Varid{third}\mathrel{=}\bot }.

The \ensuremath{\mathit{Nest}} monad can retrieve the situation, if we suspend the recursive call.
\begin{hscode}\SaveRestoreHook
\column{B}{@{}>{\hspre}l<{\hspost}@{}}%
\column{14}{@{}>{\hspre}l<{\hspost}@{}}%
\column{16}{@{}>{\hspre}l<{\hspost}@{}}%
\column{E}{@{}>{\hspre}l<{\hspost}@{}}%
\>[B]{}\Varid{thirdN}\mathbin{::}\mathit{Nest}\;\Conid{Distr}\;\Conid{Int}{}\<[E]%
\\
\>[B]{}\Varid{thirdN}\mathrel{=}\mathbf{do}\;{}\<[14]%
\>[14]{}\Varid{x}\leftarrow \Varid{lift}\;\Varid{coin}{}\<[E]%
\\
\>[14]{}\Varid{y}\leftarrow \Varid{lift}\;\Varid{coin}{}\<[E]%
\\
\>[14]{}\mathbf{case}\;(\Varid{x},\Varid{y})\;\mathbf{of}{}\<[E]%
\\
\>[14]{}\hsindent{2}{}\<[16]%
\>[16]{}(\mathrm{1},\mathrm{1})\to \Varid{return}\;\mathrm{0}{}\<[E]%
\\
\>[14]{}\hsindent{2}{}\<[16]%
\>[16]{}(\mathrm{1},\mathrm{0})\to \Varid{return}\;\mathrm{1}{}\<[E]%
\\
\>[14]{}\hsindent{2}{}\<[16]%
\>[16]{}(\mathrm{0},\mathrm{1})\to \Varid{return}\;\mathrm{2}{}\<[E]%
\\
\>[14]{}\hsindent{2}{}\<[16]%
\>[16]{}(\mathrm{0},\mathrm{0})\to \Varid{mark}\sequ \Varid{thirdN}{}\<[E]%
\ColumnHook
\end{hscode}\resethooks
What can we do with \ensuremath{\Varid{thirdN}}? One possibility is to get an ``approximation'' of the structure with the following function, which cuts the subcomputations if the recursion is deeper than the specified argument. All the computations that are too deep are replaced with \ensuremath{\Conid{Nothing}}.
\begin{hscode}\SaveRestoreHook
\column{B}{@{}>{\hspre}l<{\hspost}@{}}%
\column{3}{@{}>{\hspre}l<{\hspost}@{}}%
\column{5}{@{}>{\hspre}l<{\hspost}@{}}%
\column{18}{@{}>{\hspre}l<{\hspost}@{}}%
\column{21}{@{}>{\hspre}c<{\hspost}@{}}%
\column{21E}{@{}l@{}}%
\column{26}{@{}>{\hspre}l<{\hspost}@{}}%
\column{E}{@{}>{\hspre}l<{\hspost}@{}}%
\>[B]{}\Varid{takeN}\mathbin{::}(\Conid{Functor}\;\Varid{m},\ \Conid{Monad}\;\Varid{m})\Rightarrow \Conid{Int}\to \mathit{Nest}\;\Varid{m}\;\Varid{a}\to \mathit{Nest}\;\Varid{m}\;(\Conid{Maybe}\;\Varid{a}){}\<[E]%
\\
\>[B]{}\Varid{takeN}\;\Varid{k}\;(\mathit{Nest}\;\Varid{m})\mathrel{=}\mathit{Nest}\;(\Varid{fmap}\;(\Varid{aux}\;\Varid{k})\;\Varid{m}){}\<[E]%
\\
\>[B]{}\hsindent{3}{}\<[3]%
\>[3]{}\mathbf{where}{}\<[E]%
\\
\>[3]{}\hsindent{2}{}\<[5]%
\>[5]{}\Varid{aux}\;\mathrm{0}\;(\Conid{Wrap}\;{}\<[18]%
\>[18]{}\anonymous {}\<[21]%
\>[21]{}){}\<[21E]%
\>[26]{}\mathrel{=}\Conid{Return}\;\Conid{Nothing}{}\<[E]%
\\
\>[3]{}\hsindent{2}{}\<[5]%
\>[5]{}\Varid{aux}\;\Varid{k}\;(\Conid{Wrap}\;{}\<[18]%
\>[18]{}\Varid{m}{}\<[21]%
\>[21]{}){}\<[21E]%
\>[26]{}\mathrel{=}\Conid{Wrap}\;(\Varid{fmap}\;(\Varid{aux}\;(\Varid{k}\mathbin{-}\mathrm{1}))\;\Varid{m}){}\<[E]%
\\
\>[3]{}\hsindent{2}{}\<[5]%
\>[5]{}\Varid{aux}\;\Varid{k}\;(\Conid{Return}\;\Varid{a}){}\<[26]%
\>[26]{}\mathrel{=}\Conid{Return}\;(\Conid{Just}\;\Varid{a}){}\<[E]%
\\[\blanklineskip]%
\>[B]{}\Varid{approx}\mathbin{::}(\Conid{Functor}\;\Varid{m},\ \Conid{Monad}\;\Varid{m})\Rightarrow \Conid{Int}\to \mathit{Nest}\;\Varid{m}\;\Varid{a}\to \Varid{m}\;(\Conid{Maybe}\;\Varid{a}){}\<[E]%
\\
\>[B]{}\Varid{approx}\;\Varid{k}\mathrel{=}\Varid{drop}\mathbin{\circ}\Varid{takeN}\;\Varid{k}{}\<[E]%
\ColumnHook
\end{hscode}\resethooks
We can ask the Haskell shell:

\begin{hscode}\SaveRestoreHook
\column{B}{@{}>{\hspre}l<{\hspost}@{}}%
\column{E}{@{}>{\hspre}l<{\hspost}@{}}%
\>[B]{}\mathtt{\triangleright}\;\Varid{approx}\;\mathrm{0}\;\Varid{thirdN}{}\<[E]%
\\
\>[B]{}[\mskip1.5mu (\mathrm{0.25},\Conid{Nothing}),(\mathrm{0.25},\Conid{Just}\;\mathrm{2}),(\mathrm{0.25},\Conid{Just}\;\mathrm{1}),(\mathrm{0.25},\Conid{Just}\;\mathrm{0})\mskip1.5mu]{}\<[E]%
\\[\blanklineskip]%
\>[B]{}\mathtt{\triangleright}\;\mathbf{let}\;\Varid{simpl}\;(\Conid{Distr}\;\Varid{xs})\mathrel{=}\Conid{Distr}\;(\Varid{map}\;(\lambda \Varid{x}\to (\Varid{sum}\;[\mskip1.5mu \Varid{p}\mid (\Varid{p},\Varid{a})\leftarrow \Varid{xs},\Varid{x}\equiv\Varid{a}\mskip1.5mu],\Varid{x}))\;(\Varid{nub}\;(\Varid{fmap}\;\Varid{snd}\;\Varid{xs}))){}\<[E]%
\\
\>[B]{}\mathtt{\triangleright}\;\Varid{simpl}\;(\Varid{approx}\;\mathrm{1}\;\Varid{thirdN}){}\<[E]%
\\
\>[B]{}[\mskip1.5mu (\mathrm{0.0625},\Conid{Nothing}),(\mathrm{0.3125},\Conid{Just}\;\mathrm{2}),(\mathrm{0.3125},\Conid{Just}\;\mathrm{1}),(\mathrm{0.3125},\Conid{Just}\;\mathrm{0})\mskip1.5mu]{}\<[E]%
\\[\blanklineskip]%
\>[B]{}\mathtt{\triangleright}\;\Varid{simpl}\;(\Varid{approx}\;\mathrm{2}\;\Varid{thirdN}){}\<[E]%
\\
\>[B]{}[\mskip1.5mu (\mathrm{0.015635},\Conid{Nothing}),(\mathrm{0.32813},\Conid{Just}\;\mathrm{2}),(\mathrm{0.32813},\Conid{Just}\;\mathrm{1}),(\mathrm{0.32813},\Conid{Just}\;\mathrm{0})\mskip1.5mu]{}\<[E]%
\\[\blanklineskip]%
\>[B]{}\mathtt{\triangleright}\;\Varid{simpl}\;(\Varid{approx}\;\mathrm{10}\;\Varid{thirdN}){}\<[E]%
\\
\>[B]{}[\mskip1.5mu (2.38419\mathrm{e}{-}7,\Conid{Nothing}),(\mathrm{0.33333},\Conid{Just}\;\mathrm{2}),(\mathrm{0.33333},\Conid{Just}\;\mathrm{1}),(\mathrm{0.33333},\Conid{Just}\;\mathrm{0})\mskip1.5mu]{}\<[E]%
\ColumnHook
\end{hscode}\resethooks

\subsection{The Prolog \ensuremath{\Varid{cut}} operator}\label{sec:cut}

Prolog's \ensuremath{\Varid{cut}} operator allows one to restrict backtracking. The moment it is reached during evaluation of a predicate, it succeeds, but also discards all the possible backtrack choices created by the predicate so far. By exposing the structure of a \ensuremath{\Conid{List}} computation, we can use this effect also in Haskell. We perform a depth-first search on a rose tree of type \ensuremath{\mathit{Nest}\;[\mskip1.5mu \mskip1.5mu]}, but once we go down a level, we never go back.

\begin{hscode}\SaveRestoreHook
\column{B}{@{}>{\hspre}l<{\hspost}@{}}%
\column{3}{@{}>{\hspre}l<{\hspost}@{}}%
\column{5}{@{}>{\hspre}l<{\hspost}@{}}%
\column{20}{@{}>{\hspre}l<{\hspost}@{}}%
\column{26}{@{}>{\hspre}c<{\hspost}@{}}%
\column{26E}{@{}l@{}}%
\column{30}{@{}>{\hspre}l<{\hspost}@{}}%
\column{E}{@{}>{\hspre}l<{\hspost}@{}}%
\>[B]{}\Varid{call}\mathbin{::}\mathit{Nest}\;[\mskip1.5mu \mskip1.5mu]\;\Varid{a}\to [\mskip1.5mu \Varid{a}\mskip1.5mu]{}\<[E]%
\\
\>[B]{}\Varid{call}\;(\mathit{Nest}\;\Varid{xs})\mathrel{=}\Varid{aux}\;\Varid{xs}{}\<[E]%
\\
\>[B]{}\hsindent{3}{}\<[3]%
\>[3]{}\mathbf{where}{}\<[E]%
\\
\>[3]{}\hsindent{2}{}\<[5]%
\>[5]{}\Varid{aux}\;[\mskip1.5mu \mskip1.5mu]{}\<[30]%
\>[30]{}\mathrel{=}[\mskip1.5mu \mskip1.5mu]{}\<[E]%
\\
\>[3]{}\hsindent{2}{}\<[5]%
\>[5]{}\Varid{aux}\;(\Conid{Return}\;\Varid{a}{}\<[20]%
\>[20]{}\mathbin{:}\Varid{xs}{}\<[26]%
\>[26]{}){}\<[26E]%
\>[30]{}\mathrel{=}\Varid{a}\mathbin{:}\Varid{aux}\;\Varid{xs}{}\<[E]%
\\
\>[3]{}\hsindent{2}{}\<[5]%
\>[5]{}\Varid{aux}\;(\Conid{Wrap}\;\Varid{as}{}\<[20]%
\>[20]{}\mathbin{:}\anonymous {}\<[26]%
\>[26]{}){}\<[26E]%
\>[30]{}\mathrel{=}\Varid{aux}\;\Varid{as}{}\<[E]%
\\[\blanklineskip]%
\>[B]{}\Varid{brace}\mathbin{::}\mathit{Nest}\;[\mskip1.5mu \mskip1.5mu]\;\Varid{a}\to \mathit{Nest}\;[\mskip1.5mu \mskip1.5mu]\;\Varid{a}{}\<[E]%
\\
\>[B]{}\Varid{brace}\mathrel{=}\Varid{lift}\mathbin{\circ}\Varid{call}{}\<[E]%
\\[\blanklineskip]%
\>[B]{}\Varid{cut}\mathbin{::}\mathit{Nest}\;[\mskip1.5mu \mskip1.5mu]\;(){}\<[E]%
\\
\>[B]{}\Varid{cut}\mathrel{=}\Varid{mark}{}\<[E]%
\ColumnHook
\end{hscode}\resethooks
Consider the following example.

\begin{hscode}\SaveRestoreHook
\column{B}{@{}>{\hspre}l<{\hspost}@{}}%
\column{8}{@{}>{\hspre}l<{\hspost}@{}}%
\column{22}{@{}>{\hspre}l<{\hspost}@{}}%
\column{E}{@{}>{\hspre}l<{\hspost}@{}}%
\>[B]{}\mathtt{\triangleright}\;{}\<[8]%
\>[8]{}\Varid{call}\;(\ \mathbf{do}\;{}\<[22]%
\>[22]{}\Varid{x}\leftarrow \Varid{lift}\;[\mskip1.5mu \mathrm{4},\mathrm{7},\mathrm{13},\mathrm{9}\mskip1.5mu]{}\<[E]%
\\
\>[22]{}\Varid{y}\leftarrow \Varid{lift}\;[\mskip1.5mu \mathrm{2},\mathrm{8},\mathrm{1}\mskip1.5mu]{}\<[E]%
\\
\>[22]{}\Varid{when}\;(\Varid{x}\mathbin{+}\Varid{y}\geq \mathrm{15})\;\Varid{cut}{}\<[E]%
\\
\>[22]{}\Varid{return}\;(\Varid{x}\mathbin{+}\Varid{y})\ ){}\<[E]%
\\
\>[B]{}[\mskip1.5mu \mathrm{6},\mathrm{12},\mathrm{5},\mathrm{9},\mathrm{15}\mskip1.5mu]{}\<[E]%
\ColumnHook
\end{hscode}\resethooks
(Here, \ensuremath{\Varid{when}} is a standard Haskell function, defined by \ensuremath{\Varid{when}\;\Varid{b}\;\Varid{m}\mathrel{=}\mathbf{if}\;\Varid{b}\;\mathbf{then}\;\Varid{m}\;\mathbf{else}\;\Varid{return}\;()}.)
First, we pick \ensuremath{\mathrm{4}} from the first list, which after choices from the second list creates \ensuremath{[\mskip1.5mu \mathrm{6},\mathrm{12},\mathrm{5}\mskip1.5mu]}. Then, we pick \ensuremath{\mathrm{7}} from the first list. We cut on the second element of the second choice (because \ensuremath{\mathrm{7}\mathbin{+}\mathrm{8}\mathrel{=}\mathrm{15}}). So, all the other choices from the second list (that is, \ensuremath{\mathrm{1}}) are discarded, as well as other choices from the first list (that is, \ensuremath{\mathrm{13}} and \ensuremath{\mathrm{9}}).

We can limit the scope of \ensuremath{\Varid{cut}} by using the \ensuremath{\Varid{brace}} function. Only choices from inside of the brace are now cut.

\begin{hscode}\SaveRestoreHook
\column{B}{@{}>{\hspre}l<{\hspost}@{}}%
\column{8}{@{}>{\hspre}l<{\hspost}@{}}%
\column{22}{@{}>{\hspre}l<{\hspost}@{}}%
\column{37}{@{}>{\hspre}l<{\hspost}@{}}%
\column{E}{@{}>{\hspre}l<{\hspost}@{}}%
\>[B]{}\mathtt{\triangleright}\;{}\<[8]%
\>[8]{}\Varid{call}\;(\ \mathbf{do}\;{}\<[22]%
\>[22]{}\Varid{x}\leftarrow \Varid{lift}\;[\mskip1.5mu \mathrm{4},\mathrm{7},\mathrm{13},\mathrm{9}\mskip1.5mu]{}\<[E]%
\\
\>[22]{}\Varid{brace}\;(\ \mathbf{do}\;{}\<[37]%
\>[37]{}\Varid{y}\leftarrow \Varid{lift}\;[\mskip1.5mu \mathrm{2},\mathrm{8},\mathrm{1}\mskip1.5mu]{}\<[E]%
\\
\>[37]{}\Varid{when}\;(\Varid{x}\mathbin{+}\Varid{y}\geq \mathrm{15})\;\Varid{cut}{}\<[E]%
\\
\>[37]{}\Varid{return}\;(\Varid{x}\mathbin{+}\Varid{y})\ )\ ){}\<[E]%
\\
\>[B]{}[\mskip1.5mu \mathrm{6},\mathrm{12},\mathrm{5},\mathrm{9},\mathrm{15},\mathrm{15},\mathrm{11},\mathrm{17}\mskip1.5mu]{}\<[E]%
\ColumnHook
\end{hscode}\resethooks

Different interpretations of the \ensuremath{\mathit{Nest}} data structure enable the definition of different search strategies, such as breadth-first search. Moreover, it can even mix two different strategies in lifted and minded parts.

\subsection{Poor man's concurrency transformer, revisited}\label{sec:poorsman}

Claessen's ``poor man's'' concurrency transformer~\cite{Claessen:1999:PMC:968592.968596} adds simple concurrency capabilities to any monad. It has two flaws. The first one is that it does not respect the laws presented in Section~\ref{sec:self}. Every lifted operation is atomic, and execution can only be interrupted in between atomic actions; this means that the evaluation of \ensuremath{\Varid{lift}\;\Varid{m}_{\mathrm{1}}\sequ \Varid{lift}\;\Varid{m}_{\mathrm{2}}} can be interrupted by an action from another thread, while \ensuremath{\Varid{lift}\;(\Varid{m}_{\mathrm{1}}\sequ \Varid{m}_{\mathrm{2}})} cannot. The second flaw is that the return type of its \ensuremath{\Varid{run}} function is \ensuremath{\Varid{m}\;()}, and so it does not allow one to collect actual results of the computation. Here, we give a version of Claessen's transformer which fixes these flaws, via a conscious use of free structures.

In Claessen's monad, the user first builds a continuation, which produces an expression, which then is interpreted by the \ensuremath{\Varid{run}} function. By augmenting the type of expressions, we can skip the continuation layer. The datatype for concurrent expressions is as follows.
\begin{hscode}\SaveRestoreHook
\column{B}{@{}>{\hspre}l<{\hspost}@{}}%
\column{E}{@{}>{\hspre}l<{\hspost}@{}}%
\>[B]{}\mathbf{data}\;\Conid{Action}\;\Varid{m}\;\Varid{a}\mathrel{=}\Conid{Par}\;(\Conid{Action}\;\Varid{m}\;\Varid{a})\;(\Conid{Action}\;\Varid{m}\;\Varid{a})\mid \Conid{Act}\;(\Varid{m}\;(\Conid{Action}\;\Varid{m}\;\Varid{a}))\mid \Conid{Done}\;\Varid{a}\mid \Conid{Kill}{}\<[E]%
\ColumnHook
\end{hscode}\resethooks
Intuitively, the \ensuremath{\Conid{Par}} constructor pairs two expressions for parallel evaluation, \ensuremath{\Conid{Act}} performs a single monadic action, \ensuremath{\Conid{Done}} terminates the computation with an answer\footnote{The \ensuremath{\Conid{Kill}} constructor is called \ensuremath{\Conid{Stop}} in Claessen's datatype, and \ensuremath{\Conid{Done}} is new.}, and \ensuremath{\Conid{Kill}} terminates the computation with no answer. We treat \ensuremath{\Conid{Action}} as a term algebra with \ensuremath{\Conid{Done}} as a constructor for variables.
\begin{hscode}\SaveRestoreHook
\column{B}{@{}>{\hspre}l<{\hspost}@{}}%
\column{3}{@{}>{\hspre}l<{\hspost}@{}}%
\column{12}{@{}>{\hspre}l<{\hspost}@{}}%
\column{19}{@{}>{\hspre}l<{\hspost}@{}}%
\column{E}{@{}>{\hspre}l<{\hspost}@{}}%
\>[B]{}\mathbf{instance}\;\Conid{Functor}\;\Varid{m}\Rightarrow \Conid{Monad}\;(\Conid{Action}\;\Varid{m})\;\mathbf{where}{}\<[E]%
\\[\blanklineskip]%
\>[B]{}\hsindent{3}{}\<[3]%
\>[3]{}\Varid{return}{}\<[19]%
\>[19]{}\mathrel{=}\Conid{Done}{}\<[E]%
\\[\blanklineskip]%
\>[B]{}\hsindent{3}{}\<[3]%
\>[3]{}\Conid{Par}\;\Varid{a}\;\Varid{b}{}\<[12]%
\>[12]{}\bind \Varid{f}{}\<[19]%
\>[19]{}\mathrel{=}\Conid{Par}\;(\Varid{a}\bind \Varid{f})\;(\Varid{b}\bind \Varid{f}){}\<[E]%
\\
\>[B]{}\hsindent{3}{}\<[3]%
\>[3]{}\Conid{Act}\;\Varid{m}{}\<[12]%
\>[12]{}\bind \Varid{f}{}\<[19]%
\>[19]{}\mathrel{=}\Conid{Act}\;(\Varid{fmap}\;(\bind \Varid{f})\;\Varid{m}){}\<[E]%
\\
\>[B]{}\hsindent{3}{}\<[3]%
\>[3]{}\Conid{Done}\;\Varid{a}{}\<[12]%
\>[12]{}\bind \Varid{f}{}\<[19]%
\>[19]{}\mathrel{=}\Varid{f}\;\Varid{a}{}\<[E]%
\\
\>[B]{}\hsindent{3}{}\<[3]%
\>[3]{}\Conid{Kill}{}\<[12]%
\>[12]{}\bind \Varid{f}{}\<[19]%
\>[19]{}\mathrel{=}\Conid{Kill}{}\<[E]%
\ColumnHook
\end{hscode}\resethooks

The \ensuremath{\Conid{Action}} datatype describes a program, but does not specify which actions form atomic chunks that should not be interrupted by other operations. This task is delegated to the \ensuremath{\mathit{Nest}} transformer. The type of our concurrent monad is then \ensuremath{\mathit{Nest}\;(\Conid{Action}\;\Varid{m})\;\Varid{a}}, for a monad \ensuremath{\Varid{m}} and answer type \ensuremath{\Varid{a}}. We define a number of operations, which allow easy construction of concurrent expressions. The functions \ensuremath{\Varid{done}} and \ensuremath{\Varid{kill}} lift the appropriate constructors. A single operation can be embedded in an \ensuremath{\Conid{Action}} data structure and lifted to the concurrent monad with \ensuremath{\Varid{act}}. There are two operators for concurrency, \ensuremath{\Varid{par}} and \ensuremath{\Varid{fork}}. The former constructs a computation from two computations of the same type. The latter starts an auxiliary thread, whose final value is ignored (see the examples below).
\begin{hscode}\SaveRestoreHook
\column{B}{@{}>{\hspre}l<{\hspost}@{}}%
\column{9}{@{}>{\hspre}l<{\hspost}@{}}%
\column{E}{@{}>{\hspre}l<{\hspost}@{}}%
\>[B]{}\mathbf{type}\;\Conid{Concurrent}\;\Varid{m}\mathrel{=}\mathit{Nest}\;(\Conid{Action}\;\Varid{m}){}\<[E]%
\\[\blanklineskip]%
\>[B]{}\Varid{done}\mathbin{::}(\Conid{Monad}\;\Varid{m})\Rightarrow \Varid{a}\to \Conid{Concurrent}\;\Varid{m}\;\Varid{a}{}\<[E]%
\\
\>[B]{}\Varid{done}\mathrel{=}\Varid{lift}\mathbin{\circ}\Conid{Done}{}\<[E]%
\\[\blanklineskip]%
\>[B]{}\Varid{kill}\mathbin{::}(\Conid{Monad}\;\Varid{m})\Rightarrow \Conid{Concurrent}\;\Varid{m}\;\Varid{a}{}\<[E]%
\\
\>[B]{}\Varid{kill}\mathrel{=}\Varid{lift}\;\Conid{Kill}{}\<[E]%
\\[\blanklineskip]%
\>[B]{}\Varid{act}\mathbin{::}(\Conid{Monad}\;\Varid{m})\Rightarrow \Varid{m}\;\Varid{a}\to \Conid{Concurrent}\;\Varid{m}\;\Varid{a}{}\<[E]%
\\
\>[B]{}\Varid{act}\;\Varid{m}\mathrel{=}\Varid{lift}\;(\Conid{Act}\;(\Varid{liftM}\;\Conid{Done}\;\Varid{m})){}\<[E]%
\\[\blanklineskip]%
\>[B]{}\Varid{par}\mathbin{::}(\Conid{Monad}\;\Varid{m})\Rightarrow \Conid{Concurrent}\;\Varid{m}\;\Varid{a}\to \Conid{Concurrent}\;\Varid{m}\;\Varid{a}\to \Conid{Concurrent}\;\Varid{m}\;\Varid{a}{}\<[E]%
\\
\>[B]{}\Varid{par}\;(\mathit{Nest}\;\Varid{m}_{1})\;(\mathit{Nest}\;\Varid{m}_{2})\mathrel{=}\mathit{Nest}\;(\Conid{Par}\;(\Conid{Done}\;(\mathit{Wrap}\;\Varid{m}_{1}))\;(\Conid{Done}\;(\mathit{Wrap}\;\Varid{m}_{2}))){}\<[E]%
\\[\blanklineskip]%
\>[B]{}\Varid{fork}\mathbin{::}(\Conid{Monad}\;\Varid{m})\Rightarrow \Conid{Concurrent}\;\Varid{m}\;\Varid{b}\to \Conid{Concurrent}\;\Varid{m}\;(){}\<[E]%
\\
\>[B]{}\Varid{fork}\;\Varid{m}{}\<[9]%
\>[9]{}\mathrel{=}\Varid{par}\;(\Varid{m}\sequ \Varid{kill})\;(\Varid{act}\;(\Varid{return}\;())){}\<[E]%
\ColumnHook
\end{hscode}\resethooks

We schedule such computations with the following \ensuremath{\Varid{round}} function. We can see \ensuremath{\Conid{Done}} as a constructor which either terminates evaluation of an atomic chunk (when it is composed with \ensuremath{\mathit{Wrap}}) or the entire thread (when it is composed with \ensuremath{\mathit{Return}}).
\begin{hscode}\SaveRestoreHook
\column{B}{@{}>{\hspre}l<{\hspost}@{}}%
\column{3}{@{}>{\hspre}l<{\hspost}@{}}%
\column{18}{@{}>{\hspre}l<{\hspost}@{}}%
\column{23}{@{}>{\hspre}l<{\hspost}@{}}%
\column{E}{@{}>{\hspre}l<{\hspost}@{}}%
\>[B]{}\Varid{round}\mathbin{::}\Conid{Monad}\;\Varid{m}\Rightarrow [\mskip1.5mu \mathit{Nest}\;(\Conid{Action}\;\Varid{m})\;\Varid{x}\mskip1.5mu]\to \Varid{m}\;[\mskip1.5mu \Varid{x}\mskip1.5mu]{}\<[E]%
\\
\>[B]{}\Varid{round}\;[\mskip1.5mu \mskip1.5mu]{}\<[23]%
\>[23]{}\mathrel{=}\Varid{return}\;[\mskip1.5mu \mskip1.5mu]{}\<[E]%
\\
\>[B]{}\Varid{round}\;(\mathit{Nest}\;\Varid{w}\mathbin{:}\Varid{as}){}\<[23]%
\>[23]{}\mathrel{=}\mathbf{case}\;\Varid{w}\;\mathbf{of}{}\<[E]%
\\
\>[B]{}\hsindent{3}{}\<[3]%
\>[3]{}\Conid{Kill}{}\<[18]%
\>[18]{}\to \Varid{round}\;\Varid{as}{}\<[E]%
\\
\>[B]{}\hsindent{3}{}\<[3]%
\>[3]{}\Conid{Done}\;(\mathit{Return}\;\Varid{x}){}\<[18]%
\>[18]{}\to \mathbf{do}\;\{\mskip1.5mu \Varid{xs}\leftarrow \Varid{round}\;\Varid{as};\ \Varid{return}\;(\Varid{x}\mathbin{:}\Varid{xs})\mskip1.5mu\}{}\<[E]%
\\
\>[B]{}\hsindent{3}{}\<[3]%
\>[3]{}\Conid{Done}\;(\mathit{Wrap}\;\Varid{a}){}\<[18]%
\>[18]{}\to \Varid{round}\;(\Varid{as}\plus [\mskip1.5mu \mathit{Nest}\;\Varid{a}\mskip1.5mu]){}\<[E]%
\\
\>[B]{}\hsindent{3}{}\<[3]%
\>[3]{}\Conid{Act}\;\Varid{m}{}\<[18]%
\>[18]{}\to \mathbf{do}\;\{\mskip1.5mu \Varid{a}\leftarrow \Varid{m};\ \Varid{round}\;([\mskip1.5mu \mathit{Nest}\;\Varid{a}\mskip1.5mu]\plus \Varid{as})\mskip1.5mu\}{}\<[E]%
\\
\>[B]{}\hsindent{3}{}\<[3]%
\>[3]{}\Conid{Par}\;\Varid{a}\;\Varid{b}{}\<[18]%
\>[18]{}\to \Varid{round}\;([\mskip1.5mu \mathit{Nest}\;\Varid{b}\mskip1.5mu]\plus \Varid{as}\plus [\mskip1.5mu \mathit{Nest}\;\Varid{a}\mskip1.5mu]){}\<[E]%
\ColumnHook
\end{hscode}\resethooks

We can test our monad as follows. In the first example, we first define two expressions: \ensuremath{\Varid{cat}} writes the string \ensuremath{\text{\tt \char34 cat\char34}} five times, relinquishing control every time the operation is performed. Similarly, \ensuremath{\Varid{fish}} writes \ensuremath{\text{\tt \char34 fish\char34}} seven times.
\begin{hscode}\SaveRestoreHook
\column{B}{@{}>{\hspre}l<{\hspost}@{}}%
\column{3}{@{}>{\hspre}l<{\hspost}@{}}%
\column{E}{@{}>{\hspre}l<{\hspost}@{}}%
\>[B]{}\mathbf{instance}\;(\Conid{Monoid}\;\Varid{s})\Rightarrow \Conid{MonadWriter}\;\Varid{s}\;(\Conid{Concurrent}\;(\Conid{Writer}\;\Varid{s}))\;\mathbf{where}{}\<[E]%
\\
\>[B]{}\hsindent{3}{}\<[3]%
\>[3]{}\Varid{tell}\mathrel{=}\Varid{act}\mathbin{\circ}\Varid{tell}{}\<[E]%
\\[\blanklineskip]%
\>[B]{}\Varid{cat}\mathbin{::}\Conid{Concurrent}\;(\Conid{Writer}\;\Conid{String})\;\Conid{Int}{}\<[E]%
\\
\>[B]{}\Varid{cat}\mathrel{=}\Varid{replicateM}\;\mathrm{5}\;(\Varid{tell}\;\text{\tt \char34 cat\char34}\sequ \Varid{mark})\sequ \Varid{return}\;\mathrm{1}{}\<[E]%
\\[\blanklineskip]%
\>[B]{}\Varid{fish}\mathbin{::}\Conid{Concurrent}\;(\Conid{Writer}\;\Conid{String})\;\Conid{Int}{}\<[E]%
\\
\>[B]{}\Varid{fish}\mathrel{=}\Varid{replicateM}\;\mathrm{7}\;(\Varid{tell}\;\text{\tt \char34 fish\char34}\sequ \Varid{mark})\sequ \Varid{return}\;\mathrm{2}{}\<[E]%
\ColumnHook
\end{hscode}\resethooks
We can test them, by running them in parallel.
\begin{hscode}\SaveRestoreHook
\column{B}{@{}>{\hspre}l<{\hspost}@{}}%
\column{7}{@{}>{\hspre}l<{\hspost}@{}}%
\column{18}{@{}>{\hspre}l<{\hspost}@{}}%
\column{E}{@{}>{\hspre}l<{\hspost}@{}}%
\>[B]{}\mathtt{\triangleright}\;{}\<[7]%
\>[7]{}\Varid{round}\;[\mskip1.5mu \mathbf{do}\;{}\<[18]%
\>[18]{}\Varid{x}\leftarrow \Varid{fish}\mathbin{`\Varid{par}`}\Varid{cat}{}\<[E]%
\\
\>[18]{}\Varid{tell}\;\text{\tt \char34 dog\char34}{}\<[E]%
\\
\>[18]{}\Varid{return}\;\Varid{x}\mskip1.5mu]{}\<[E]%
\\
\>[B]{}(\text{\tt \char34 catfishcatfishcatfishcatfishcatfishdogfishfishdog\char34},[\mskip1.5mu \mathrm{1},\mathrm{2}\mskip1.5mu]){}\<[E]%
\ColumnHook
\end{hscode}\resethooks
The results of all parallel threads called with \ensuremath{\Varid{par}}, in this example \ensuremath{[\mskip1.5mu \mathrm{1},\mathrm{2}\mskip1.5mu]}, are returned in a list. The operation \ensuremath{\Varid{tell}\;\text{\tt \char34 dog\char34}} is bound to both threads.
 
We can now run \ensuremath{\Varid{fish}} in a separate, auxiliary thread. The thread is run on the side, the following actions are not bound to it, and its result is not returned with the overall result.
\begin{hscode}\SaveRestoreHook
\column{B}{@{}>{\hspre}l<{\hspost}@{}}%
\column{7}{@{}>{\hspre}l<{\hspost}@{}}%
\column{18}{@{}>{\hspre}l<{\hspost}@{}}%
\column{E}{@{}>{\hspre}l<{\hspost}@{}}%
\>[B]{}\mathtt{\triangleright}\;{}\<[7]%
\>[7]{}\Varid{round}\;[\mskip1.5mu \mathbf{do}\;{}\<[18]%
\>[18]{}\Varid{fork}\;\Varid{fish}{}\<[E]%
\\
\>[18]{}\Varid{x}\leftarrow \Varid{cat}{}\<[E]%
\\
\>[18]{}\Varid{tell}\;\text{\tt \char34 dog\char34}{}\<[E]%
\\
\>[18]{}\Varid{return}\;\Varid{x}\mskip1.5mu]{}\<[E]%
\\
\>[B]{}(\text{\tt \char34 catfishcatfishcatfishcatfishcatfishdogfishfish\char34},[\mskip1.5mu \mathrm{1}\mskip1.5mu]){}\<[E]%
\ColumnHook
\end{hscode}\resethooks

We can also define a version of \ensuremath{\Varid{fish}} that is performed atomically.

\begin{hscode}\SaveRestoreHook
\column{B}{@{}>{\hspre}l<{\hspost}@{}}%
\column{E}{@{}>{\hspre}l<{\hspost}@{}}%
\>[B]{}\Varid{fish'}\mathbin{::}\Conid{Concurrent}\;(\Conid{Writer}\;\Conid{String})\;\Conid{Int}{}\<[E]%
\\
\>[B]{}\Varid{fish'}\mathrel{=}\Varid{replicateM}\;\mathrm{7}\;(\Varid{tell}\;\text{\tt \char34 fish\char34})\sequ \Varid{return}\;\mathrm{2}{}\<[E]%
\ColumnHook
\end{hscode}\resethooks
We can see that \ensuremath{\Varid{round}} does not separate the calls of \ensuremath{\Varid{tell}\;\text{\tt \char34 fish\char34}}.
\begin{hscode}\SaveRestoreHook
\column{B}{@{}>{\hspre}l<{\hspost}@{}}%
\column{7}{@{}>{\hspre}l<{\hspost}@{}}%
\column{18}{@{}>{\hspre}l<{\hspost}@{}}%
\column{E}{@{}>{\hspre}l<{\hspost}@{}}%
\>[B]{}\mathtt{\triangleright}\;{}\<[7]%
\>[7]{}\Varid{round}\;[\mskip1.5mu \mathbf{do}\;{}\<[18]%
\>[18]{}\Varid{fork}\;\Varid{fish'}{}\<[E]%
\\
\>[18]{}\Varid{x}\leftarrow \Varid{cat}{}\<[E]%
\\
\>[18]{}\Varid{tell}\;\text{\tt \char34 dog\char34}{}\<[E]%
\\
\>[18]{}\Varid{return}\;\Varid{x}\mskip1.5mu]{}\<[E]%
\\
\>[B]{}(\text{\tt \char34 catfishfishfishfishfishfishfishcatcatcatcatdog\char34},[\mskip1.5mu \mathrm{1}\mskip1.5mu]){}\<[E]%
\ColumnHook
\end{hscode}\resethooks

\section{Related work} \label{sec:related}

The idea of separation of syntax and semantics of monadic computations is not new. It is the very foundation of the success of monads as a tool for encapsulating impure behaviour in pure languages~\cite{PeytonJones:1993:IFP:158511.158524}---for example, the way Haskell integrates impure effects such as I/O within a pure language is to make the pure evaluation \emph{construct} a syntactic term, which is subsequently \emph{interpreted} by the run-time system.

The work most related to ours is the Unimo framework introduced by Lin~\cite{Lin:2006:PMO:1159803.1159840}, which is an embedded domain-specific language designed to modularise construction of monads in Haskell. Even though Lin's motivation and toolbox significantly differ from ours, he came to the same conclusion that exposing the structure of computations allows more functionality to be added to existing monads. We share a strong flavour of aspect-oriented programming.

There is a resemblance between resumptions, used for modelling semantics of concurrency~\cite{Ganz:1999:TS:317636.317779, DBLP:conf/mfcs/HennessyP79}, and the \ensuremath{\Conid{Nest}} monad. A \emph{resumption monad transformer} is used, for example, by Papaspyrou to model semantics of concurrency in domain theory~\cite{citeulike:155438}, and by Harrison in his ``cheap'' concurrency~\cite{citeulike:276362, conf/other/Harrison06}. They both use a definition similar to ours from Section~\ref{sec:free} (though they fail to mention the connection between free monads and resumptions), and as a result their constructions are not transformers in the sense of Section~\ref{sec:self}.

Harrison~\cite{citeulike:276362} is not quite right in claiming that Claessen's monad is based on first-class continuations. A version of the continuation monad is used only to build a syntactic term, which serves as the backbone of the concurrent computation. The term is not a monad, since it lacks free variables, but it reveals the structure of resumptions (notice the type of the constructor \ensuremath{\Conid{Atom}\mathbin{::}\Varid{m}\;(\Conid{Action}\;\Varid{m})\to \Conid{Action}\;\Varid{m}}). Similarly, in Swierstra and Altenkirch's functional specification of concurrency~\cite{Swierstra:2007:BB:1291201.1291206}, resumptions are used implicitly, and control is surrendered by a thread whenever it wants to communicate with other parts of the concurrent system (which is denoted by a constructor of the free structure \ensuremath{\Conid{IO}_{\Varid{c}}}). As we show in Sections~\ref{sec:partiality} and~\ref{sec:poorsman}, the \ensuremath{\mathit{Nest}} monad transformer allows one to separate the concepts of composition of computations and yielding, by an explicit use of \ensuremath{\Varid{mark}}.

A definition of a resumption transformer, which satisfies the laws from Section~\ref{sec:self}, and is in fact isomorphic to \ensuremath{\Conid{Nest}}, was already given by Cenciarelli and Moggi~\cite{Cenciarelli93asyntactic}, but practical applications in programming were not studied. Also---as pointed to us by the anonymous reviewers---the \ensuremath{\Conid{Nest}} transformer can be obtained via Hyland, Plotkin and Power's \emph{sum} construction~\cite{Hyland:2006:CES:1161493.1161499}, as a sum of a monad and an \ensuremath{\Conid{Identity}}-generated free monad. Though Hyland \emph{et al}.'s construction provides a simpler proof that \ensuremath{\Conid{Nest}} can be given a monad structure (using a distributive law between a monad \ensuremath{\Varid{m}} and the \ensuremath{\Varid{m}}-generated free monad), our definition of \ensuremath{\Varid{join}} is not intensionally similar to the one arising from the \ensuremath{\Varid{sum}} construction, and issues like efficiency should also be taken into account. A naive implementation of \ensuremath{\Varid{join}} in the sum construction traverses the structure twice (once to apply the distributive law, and once to join the free structure), while \ensuremath{\Varid{join}} for \ensuremath{\Conid{Nest}} needs to traverse the structure only once. On the other hand, the \ensuremath{\Varid{sum}} construction allows one to include an additional functor---the datatype in question is of the form \ensuremath{\Conid{M}\;(\Conid{Free}\;(\Conid{F}\mathbin{\circ}\Conid{M})\;\Varid{a})}---which may help to generalize our notion of tracing in the future.

The interleaving between pure data and monadic structure was also considered by Filinski and St{\o}vring~\cite{Filinski:2007:IRE:1291151.1291168}, and in forthcoming work by Atkey \emph{et al.}~\cite{induction-with-effects}. They give proof principles for reasoning about datatypes that include effects, for example a stream in which tails are always guarded by I/O actions.

\section{Future work}\label{sec:future}

So far, we failed to mention the mother and father of all purely functional monads, that is the continuation and state monads. We do not have much to say about continuations, but we see a lot of applications in tracing the \ensuremath{\Conid{State}} monad.

\paragraph{Functional specifications of effects.} The idea behind functional specifications of effects in pure languages is to model the logic of an effectful construct in the pure core of the language. For example, such a specification may consist of a datatype representing a model of the outside world and a variation of the \ensuremath{\Conid{State}} monad whose state modifications mirror the actions of the side-effecting monad. This way, we can translate a program into its pure equivalent, test it, and reason about it no differently from how we would reason about any other pure program.

The existing frameworks for specifying ``effectful'' Haskell in ``pure'' Haskell---like those proposed by Swierstra and Altenkirch~\cite{swierstra:thesis, Swierstra:2007:BB:1291201.1291206} and Butterfield \emph{et al}.~\cite{Butterfield:2001:PCP:647980.743391, Dowse:2002:PMC:1756972.1756977}---do not concentrate on non-terminating computations, which are of little use in the pure world, but which are back in the spotlight in the presence of effects like I/O and concurrency. For example, Butterfield \emph{et al.} model the interaction between programs and a filesystem by means of the \ensuremath{\Conid{State}} monad, which for an initial state (of the filesystem) produces a final value and a final state. In case of non-terminating programs, no final state exists, so the whole model becomes useless. What we are really interested in is not a final state, but the whole (possibly infinite) sequence of subsequent states of the filesystem, or a trace of all the interactions. In such a setting, we can use coinduction as a reasoning tool, which coincides with the non-strict semantics of Haskell, in which we try to embed our model.

As mentioned before, our approach can transform monolithic computations into coinductive unfolding of traces. That is why we propose to use a different monad as a basis for functional specifications. It is a monad which produces not only the final state, but the whole (possibly infinite) stream of intermediate states.

\begin{hscode}\SaveRestoreHook
\column{B}{@{}>{\hspre}l<{\hspost}@{}}%
\column{9}{@{}>{\hspre}l<{\hspost}@{}}%
\column{10}{@{}>{\hspre}l<{\hspost}@{}}%
\column{20}{@{}>{\hspre}l<{\hspost}@{}}%
\column{22}{@{}>{\hspre}l<{\hspost}@{}}%
\column{E}{@{}>{\hspre}l<{\hspost}@{}}%
\>[B]{}\mathbf{data}\;{}\<[9]%
\>[9]{}\Conid{Trace}\;\Varid{s}\;\Varid{a}{}\<[20]%
\>[20]{}\mathrel{=}\Conid{TCons}\;\Varid{s}\;(\Conid{Trace}\;\Varid{s}\;\Varid{a})\mid \Conid{Nil}\;\Varid{a}{}\<[E]%
\\[\blanklineskip]%
\>[B]{}\mathbf{newtype}\;{}\<[10]%
\>[10]{}\Conid{States}\;\Varid{s}\;\Varid{a}{}\<[22]%
\>[22]{}\mathrel{=}\Conid{States}\{\mskip1.5mu \Varid{runStates}\mathbin{::}\Varid{s}\to \Conid{Trace}\;\Varid{s}\;\Varid{a}\mskip1.5mu\}{}\<[E]%
\ColumnHook
\end{hscode}\resethooks
We leave the exact implementation of instances of \ensuremath{\Conid{Monad}\;(\Conid{States}\;\Varid{s})}, \ensuremath{\Conid{MonadTrans}\;(\Conid{State}\;\Varid{s})\;(\Conid{States}\;\Varid{s})} and \ensuremath{\Conid{MonadTrace}\;(\Conid{States}\;\Varid{s})} to the reader as an exercise. The \ensuremath{\Varid{mark}} operation should accumulate the current state in the \ensuremath{\Conid{Trace}}.

What is the relationship between \ensuremath{\Conid{States}\;\Varid{s}} and \ensuremath{\mathit{Nest}\;(\Conid{State}\;\Varid{s})}? It is possible to interpret free parts of \ensuremath{\mathit{Nest}\;(\Conid{State}\;\Varid{s})} in a suitable way, that is to define a monad morphism of type \ensuremath{\mathit{Nest}\;(\Conid{State}\;\Varid{s})\to \Conid{States}\;\Varid{s}}. We can also suspect a different kind of generality, since \ensuremath{\Conid{State}} is a composition of two adjoint functors, namely \ensuremath{\Conid{State}\;\Varid{s}\mathrel{=}\Conid{Reader}\;\Varid{s}\mathbin{\circ}\Conid{Writer}\;\Varid{s}}, while \ensuremath{\Conid{States}\;\Varid{s}\mathrel{=}\Conid{Reader}\;\Varid{s}\mathbin{\circ}\Conid{Free}\;(\Conid{Writer}\;\Varid{s})}.


\bibliographystyle{eptcs}
\bibliography{tr}

\appendix
\def\thesection{Appendix~\Alph{section}}

\newpage
\section{Proof that \ensuremath{\mathit{Nest}} is a monad}\label{sec:self-monad-proof}
We prove the properties (2) and (3) from Section~\ref{sec:self-monad} by induction, assuming an initial-algebra reading of the datatype \ensuremath{\Conid{Free}\;\Varid{m}}---that is, we assume a well-founded ordering on the subterms of any value of type \ensuremath{\Conid{Free}\;\Varid{m}\;\Varid{a}}. (We have to resort to something like induction, because the definition of \ensuremath{\Varid{prod}} above isn't in the form of a standard recurson pattern---in particular, it is not a fold.) For brevity, we omit the \ensuremath{\mathit{Nest}} constructor.

Property (2): The case for \ensuremath{\Conid{Return}} is straightforward. For the \ensuremath{\Conid{Wrap}} case, we assume that the property holds for each element of data structure~\ensuremath{\Varid{m}}; that is, that
\begin{hscode}\SaveRestoreHook
\column{B}{@{}>{\hspre}l<{\hspost}@{}}%
\column{E}{@{}>{\hspre}l<{\hspost}@{}}%
\>[B]{}\Conid{M}\;(\Varid{prod}\mathbin{\circ}\Conid{F}\;\Varid{return})\;\Varid{m}\mathrel{=}\Conid{M}\;\Varid{return}_{\Conid{M}}\;\Varid{m}{}\<[E]%
\ColumnHook
\end{hscode}\resethooks
Then we calculate:
\begin{hscode}\SaveRestoreHook
\column{B}{@{}>{\hspre}c<{\hspost}@{}}%
\column{BE}{@{}l@{}}%
\column{3}{@{}>{\hspre}l<{\hspost}@{}}%
\column{5}{@{}>{\hspre}l<{\hspost}@{}}%
\column{E}{@{}>{\hspre}l<{\hspost}@{}}%
\>[3]{}(\Varid{prod}\mathbin{\circ}\Conid{F}\;\Varid{return})\;(\Conid{Wrap}\;\Varid{m}){}\<[E]%
\\
\>[B]{}\mathrel{=}{}\<[BE]%
\>[5]{}\mbox{\commentbegin  naturality of \ensuremath{\Conid{Wrap}\mathbin{::}\Conid{M}\;\Conid{F}\to \Conid{F}}  \commentend}{}\<[E]%
\\
\>[B]{}\hsindent{3}{}\<[3]%
\>[3]{}(\Varid{prod}\mathbin{\circ}\Conid{Wrap}\mathbin{\circ}\Conid{M}\;\Conid{F}\;\Varid{return})\;\Varid{m}{}\<[E]%
\\
\>[B]{}\mathrel{=}{}\<[BE]%
\>[5]{}\mbox{\commentbegin  definition of \ensuremath{\Varid{prod}}  \commentend}{}\<[E]%
\\
\>[B]{}\hsindent{3}{}\<[3]%
\>[3]{}(\Varid{return}_{\Conid{M}}\mathbin{\circ}\Conid{Wrap}\mathbin{\circ}\Varid{join}_{\Conid{M}}\mathbin{\circ}\Conid{M}\;\Varid{prod}\mathbin{\circ}\Conid{M}\;\Conid{F}\;\Varid{return})\;\Varid{m}{}\<[E]%
\\
\>[B]{}\mathrel{=}{}\<[BE]%
\>[5]{}\mbox{\commentbegin  functors  \commentend}{}\<[E]%
\\
\>[B]{}\hsindent{3}{}\<[3]%
\>[3]{}(\Varid{return}_{\Conid{M}}\mathbin{\circ}\Conid{Wrap}\mathbin{\circ}\Varid{join}_{\Conid{M}}\mathbin{\circ}\Conid{M}\;(\Varid{prod}\mathbin{\circ}\Conid{F}\;\Varid{return}))\;\Varid{m}{}\<[E]%
\\
\>[B]{}\mathrel{=}{}\<[BE]%
\>[5]{}\mbox{\commentbegin  induction  \commentend}{}\<[E]%
\\
\>[B]{}\hsindent{3}{}\<[3]%
\>[3]{}(\Varid{return}_{\Conid{M}}\mathbin{\circ}\Conid{Wrap}\mathbin{\circ}\Varid{join}_{\Conid{M}}\mathbin{\circ}\Conid{M}\;\Varid{return}_{\Conid{M}})\;\Varid{m}{}\<[E]%
\\
\>[B]{}\mathrel{=}{}\<[BE]%
\>[5]{}\mbox{\commentbegin  \ensuremath{\Conid{M}} as a monad  \commentend}{}\<[E]%
\\
\>[B]{}\hsindent{3}{}\<[3]%
\>[3]{}(\Varid{return}_{\Conid{M}}\mathbin{\circ}\Conid{Wrap})\;\Varid{m}{}\<[E]%
\ColumnHook
\end{hscode}\resethooks

Property (3): The case for \ensuremath{\Conid{Return}} is again straightforward. In the \ensuremath{\Conid{Wrap}} case, we again assume that the property holds for each element of~\ensuremath{\Varid{m}}:
\begin{hscode}\SaveRestoreHook
\column{B}{@{}>{\hspre}l<{\hspost}@{}}%
\column{E}{@{}>{\hspre}l<{\hspost}@{}}%
\>[B]{}\Conid{M}\;(\Varid{join}\mathbin{\circ}\Varid{prod})\;\Varid{m}\mathrel{=}\Conid{M}\;(\Varid{prod}\mathbin{\circ}\Conid{F}\;\Varid{join})\;\Varid{m}{}\<[E]%
\ColumnHook
\end{hscode}\resethooks
Then we calculate:
\begin{hscode}\SaveRestoreHook
\column{B}{@{}>{\hspre}c<{\hspost}@{}}%
\column{BE}{@{}l@{}}%
\column{3}{@{}>{\hspre}l<{\hspost}@{}}%
\column{4}{@{}>{\hspre}l<{\hspost}@{}}%
\column{E}{@{}>{\hspre}l<{\hspost}@{}}%
\>[3]{}(\Varid{join}\mathbin{\circ}\Varid{prod})\;(\Conid{Wrap}\;\Varid{m}){}\<[E]%
\\
\>[B]{}\mathrel{=}{}\<[BE]%
\>[4]{}\mbox{\commentbegin  definition of \ensuremath{\Varid{join}}  \commentend}{}\<[E]%
\\
\>[B]{}\hsindent{3}{}\<[3]%
\>[3]{}(\Varid{join}_{\Conid{M}}\mathbin{\circ}\Conid{M}\;\Varid{prod}\mathbin{\circ}\Varid{prod}\mathbin{\circ}\Conid{Wrap})\;\Varid{m}{}\<[E]%
\\
\>[B]{}\mathrel{=}{}\<[BE]%
\>[4]{}\mbox{\commentbegin  definition of \ensuremath{\Varid{prod}}  \commentend}{}\<[E]%
\\
\>[B]{}\hsindent{3}{}\<[3]%
\>[3]{}(\Varid{join}_{\Conid{M}}\mathbin{\circ}\Conid{M}\;\Varid{prod}\mathbin{\circ}\Varid{return}_{\Conid{M}}\mathbin{\circ}\Conid{Wrap}\mathbin{\circ}\Varid{join}_{\Conid{M}}\mathbin{\circ}\Conid{M}\;\Varid{prod})\;\Varid{m}{}\<[E]%
\\
\>[B]{}\mathrel{=}{}\<[BE]%
\>[4]{}\mbox{\commentbegin  naturality of \ensuremath{\Varid{return}_{\Conid{M}}}  \commentend}{}\<[E]%
\\
\>[B]{}\hsindent{3}{}\<[3]%
\>[3]{}(\Varid{join}_{\Conid{M}}\mathbin{\circ}\Varid{return}_{\Conid{M}}\mathbin{\circ}\Varid{prod}\mathbin{\circ}\Conid{Wrap}\mathbin{\circ}\Varid{join}_{\Conid{M}}\mathbin{\circ}\Conid{M}\;\Varid{prod})\;\Varid{m}{}\<[E]%
\\
\>[B]{}\mathrel{=}{}\<[BE]%
\>[4]{}\mbox{\commentbegin  \ensuremath{\Conid{M}} as a monad  \commentend}{}\<[E]%
\\
\>[B]{}\hsindent{3}{}\<[3]%
\>[3]{}(\Varid{prod}\mathbin{\circ}\Conid{Wrap}\mathbin{\circ}\Varid{join}_{\Conid{M}}\mathbin{\circ}\Conid{M}\;\Varid{prod})\;\Varid{m}{}\<[E]%
\\
\>[B]{}\mathrel{=}{}\<[BE]%
\>[4]{}\mbox{\commentbegin  definition of \ensuremath{\Varid{prod}}  \commentend}{}\<[E]%
\\
\>[B]{}\hsindent{3}{}\<[3]%
\>[3]{}(\Varid{return}_{\Conid{M}}\mathbin{\circ}\Conid{Wrap}\mathbin{\circ}\Varid{join}_{\Conid{M}}\mathbin{\circ}\Conid{M}\;\Varid{prod}\mathbin{\circ}\Varid{join}_{\Conid{M}}\mathbin{\circ}\Conid{M}\;\Varid{prod})\;\Varid{m}{}\<[E]%
\\
\>[B]{}\mathrel{=}{}\<[BE]%
\>[4]{}\mbox{\commentbegin  naturality of \ensuremath{\Varid{join}_{\Conid{M}}}  \commentend}{}\<[E]%
\\
\>[B]{}\hsindent{3}{}\<[3]%
\>[3]{}(\Varid{return}_{\Conid{M}}\mathbin{\circ}\Conid{Wrap}\mathbin{\circ}\Varid{join}_{\Conid{M}}\mathbin{\circ}\Varid{join}_{\Conid{M}}\mathbin{\circ}\Conid{M}\;\Conid{M}\;\Varid{prod}\mathbin{\circ}\Conid{M}\;\Varid{prod})\;\Varid{m}{}\<[E]%
\\
\>[B]{}\mathrel{=}{}\<[BE]%
\>[4]{}\mbox{\commentbegin  \ensuremath{\Conid{M}} as a monad  \commentend}{}\<[E]%
\\
\>[B]{}\hsindent{3}{}\<[3]%
\>[3]{}(\Varid{return}_{\Conid{M}}\mathbin{\circ}\Conid{Wrap}\mathbin{\circ}\Varid{join}_{\Conid{M}}\mathbin{\circ}\Conid{M}\;\Varid{join}_{\Conid{M}}\mathbin{\circ}\Conid{M}\;\Conid{M}\;\Varid{prod}\mathbin{\circ}\Conid{M}\;\Varid{prod})\;\Varid{m}{}\<[E]%
\\
\>[B]{}\mathrel{=}{}\<[BE]%
\>[4]{}\mbox{\commentbegin  functors  \commentend}{}\<[E]%
\\
\>[B]{}\hsindent{3}{}\<[3]%
\>[3]{}(\Varid{return}_{\Conid{M}}\mathbin{\circ}\Conid{Wrap}\mathbin{\circ}\Varid{join}_{\Conid{M}}\mathbin{\circ}\Conid{M}\;(\Varid{join}_{\Conid{M}}\mathbin{\circ}\Conid{M}\;\Varid{prod}\mathbin{\circ}\Varid{prod}))\;\Varid{m}{}\<[E]%
\\
\>[B]{}\mathrel{=}{}\<[BE]%
\>[4]{}\mbox{\commentbegin  definition of \ensuremath{\Varid{join}}; induction  \commentend}{}\<[E]%
\\
\>[B]{}\hsindent{3}{}\<[3]%
\>[3]{}(\Varid{return}_{\Conid{M}}\mathbin{\circ}\Conid{Wrap}\mathbin{\circ}\Varid{join}_{\Conid{M}}\mathbin{\circ}\Conid{M}\;(\Varid{prod}\mathbin{\circ}\Conid{F}\;\Varid{join}))\;\Varid{m}{}\<[E]%
\\
\>[B]{}\mathrel{=}{}\<[BE]%
\>[4]{}\mbox{\commentbegin  functors  \commentend}{}\<[E]%
\\
\>[B]{}\hsindent{3}{}\<[3]%
\>[3]{}(\Varid{return}_{\Conid{M}}\mathbin{\circ}\Conid{Wrap}\mathbin{\circ}\Varid{join}_{\Conid{M}}\mathbin{\circ}\Conid{M}\;\Varid{prod}\mathbin{\circ}\Conid{M}\;\Conid{F}\;\Varid{join})\;\Varid{m}{}\<[E]%
\\
\>[B]{}\mathrel{=}{}\<[BE]%
\>[4]{}\mbox{\commentbegin  definition of \ensuremath{\Varid{join}}  \commentend}{}\<[E]%
\\
\>[B]{}\hsindent{3}{}\<[3]%
\>[3]{}(\Varid{return}_{\Conid{M}}\mathbin{\circ}\Conid{Wrap}\mathbin{\circ}\Varid{join}\mathbin{\circ}\Conid{M}\;\Conid{F}\;\Varid{join})\;\Varid{m}{}\<[E]%
\\
\>[B]{}\mathrel{=}{}\<[BE]%
\>[4]{}\mbox{\commentbegin  definition of \ensuremath{\Varid{prod}}  \commentend}{}\<[E]%
\\
\>[B]{}\hsindent{3}{}\<[3]%
\>[3]{}(\Varid{prod}\mathbin{\circ}\Conid{Wrap}\mathbin{\circ}\Conid{M}\;\Conid{F}\;\Varid{join})\;\Varid{m}{}\<[E]%
\\
\>[B]{}\mathrel{=}{}\<[BE]%
\>[4]{}\mbox{\commentbegin  naturality of \ensuremath{\Conid{Wrap}\mathbin{::}\Conid{M}\;\Conid{F}\stackrel{.}{\rightarrow}\Conid{F}}  \commentend}{}\<[E]%
\\
\>[B]{}\hsindent{3}{}\<[3]%
\>[3]{}(\Varid{prod}\mathbin{\circ}\Conid{F}\;\Varid{join})\;(\Conid{Wrap}\;\Varid{m}){}\<[E]%
\ColumnHook
\end{hscode}\resethooks

\section{Proof that \ensuremath{\mathit{Nest}} is a tracer}\label{sec:appendix}

Here, we prove that \ensuremath{\mathit{Nest}} is a tracer (see Sections~\ref{sec:tracer} and~\ref{sec:nest-tracer} for the definitions). The equalities \ensuremath{\Varid{drop}\mathbin{\circ}\Varid{lift}\mathrel{=}\Varid{id}} and \ensuremath{\Varid{drop}\;\Varid{mark}\mathrel{=}\Varid{return}\;()} are straightforward.

We prove the equality \ensuremath{\Varid{lift}\mathbin{\circ}\mathit{return_M}\mathrel{=}\mathit{return_N}} as follows.

\begin{hscode}\SaveRestoreHook
\column{B}{@{}>{\hspre}l<{\hspost}@{}}%
\column{3}{@{}>{\hspre}l<{\hspost}@{}}%
\column{4}{@{}>{\hspre}l<{\hspost}@{}}%
\column{E}{@{}>{\hspre}l<{\hspost}@{}}%
\>[3]{}\Varid{lift}\mathbin{\circ}\mathit{return_M}{}\<[E]%
\\
\>[B]{}\mathrel{=}{}\<[4]%
\>[4]{}\mbox{\commentbegin  definition of \ensuremath{\Varid{lift}}  \commentend}{}\<[E]%
\\
\>[B]{}\hsindent{3}{}\<[3]%
\>[3]{}\Conid{M}\;\Conid{Return}\mathbin{\circ}\mathit{return_M}{}\<[E]%
\\
\>[B]{}\mathrel{=}\mbox{\commentbegin  naturality of \ensuremath{\mathit{return_M}}  \commentend}{}\<[E]%
\\
\>[B]{}\hsindent{3}{}\<[3]%
\>[3]{}\mathit{return_M}\mathbin{\circ}\Conid{Return}{}\<[E]%
\\
\>[B]{}\mathrel{=}\mbox{\commentbegin  definition of \ensuremath{\mathit{return_N}}  \commentend}{}\<[E]%
\\
\>[B]{}\hsindent{3}{}\<[3]%
\>[3]{}\mathit{return_N}{}\<[E]%
\ColumnHook
\end{hscode}\resethooks

The fact that \ensuremath{\Varid{lift}\;\Varid{c}\;\bind_N\;\Varid{lift}\mathbin{\circ}\Varid{f}\mathrel{=}\Varid{lift}\;(\Varid{c}\;\bind_M\;\Varid{f})} follows from the following.

\def\commentbegin{\quad\{\ }
\def\commentend{\}}
\begin{hscode}\SaveRestoreHook
\column{B}{@{}>{\hspre}c<{\hspost}@{}}%
\column{BE}{@{}l@{}}%
\column{3}{@{}>{\hspre}l<{\hspost}@{}}%
\column{5}{@{}>{\hspre}l<{\hspost}@{}}%
\column{E}{@{}>{\hspre}l<{\hspost}@{}}%
\>[3]{}\Varid{lift}\;\Varid{c}\;\bind_N\;\Varid{lift}\mathbin{\circ}\Varid{f}{}\<[E]%
\\
\>[B]{}={}\<[BE]%
\>[5]{}\mbox{\commentbegin  definition of \ensuremath{\bind_N}  \commentend}{}\<[E]%
\\
\>[B]{}\hsindent{3}{}\<[3]%
\>[3]{}\Varid{join}_{\Conid{N}}\;(N\;(\Varid{lift}\mathbin{\circ}\Varid{f})\;(\Varid{lift}\;\Varid{c})){}\<[E]%
\\
\>[B]{}={}\<[BE]%
\>[5]{}\mbox{\commentbegin  definition of \ensuremath{\Varid{lift}}  \commentend}{}\<[E]%
\\
\>[B]{}\hsindent{3}{}\<[3]%
\>[3]{}\Varid{join}_{\Conid{N}}\;(N\;(\Conid{M}\;\mathit{Return}\mathbin{\circ}\Varid{f})\;(\Conid{M}\;\mathit{Return}\;\Varid{c})){}\<[E]%
\\
\>[B]{}={}\<[BE]%
\>[5]{}\mbox{\commentbegin  definition of \ensuremath{N}  \commentend}{}\<[E]%
\\
\>[B]{}\hsindent{3}{}\<[3]%
\>[3]{}\Varid{join}_{\Conid{N}}\;(\Conid{MF}\;(\Conid{M}\;\mathit{Return}\mathbin{\circ}\Varid{f})\;(\Conid{M}\;\mathit{Return}\;\Varid{c})){}\<[E]%
\\
\>[B]{}={}\<[BE]%
\>[5]{}\mbox{\commentbegin  definition of \ensuremath{\Varid{join}_{\Conid{N}}}  \commentend}{}\<[E]%
\\
\>[B]{}\hsindent{3}{}\<[3]%
\>[3]{}\mathit{join_M}\;(\Conid{M}\;\Varid{prod}\;(\Conid{MF}\;(\Conid{M}\;\mathit{Return}\mathbin{\circ}\Varid{f})\;(\Conid{M}\;\mathit{Return}\;\Varid{c}))){}\<[E]%
\\
\>[B]{}={}\<[BE]%
\>[5]{}\mbox{\commentbegin  functor  \commentend}{}\<[E]%
\\
\>[B]{}\hsindent{3}{}\<[3]%
\>[3]{}\mathit{join_M}\;(\Conid{M}\;(\Varid{prod}\mathbin{\circ}\Conid{F}\;(\Conid{M}\;\mathit{Return}\mathbin{\circ}\Varid{f})\mathbin{\circ}\mathit{Return})\;\Varid{c}){}\<[E]%
\\
\>[B]{}={}\<[BE]%
\>[5]{}\mbox{\commentbegin  definition of \ensuremath{\Conid{F}}  \commentend}{}\<[E]%
\\
\>[B]{}\hsindent{3}{}\<[3]%
\>[3]{}\mathit{join_M}\;(\Conid{M}\;(\Varid{prod}\mathbin{\circ}\mathit{Return}\mathbin{\circ}\Conid{M}\;\mathit{Return}\mathbin{\circ}\Varid{f})\;\Varid{c}){}\<[E]%
\\
\>[B]{}={}\<[BE]%
\>[5]{}\mbox{\commentbegin  definition of \ensuremath{\mathit{prod}}  \commentend}{}\<[E]%
\\
\>[B]{}\hsindent{3}{}\<[3]%
\>[3]{}\mathit{join_M}\;(\Conid{M}\;(\Conid{M}\;\mathit{Return}\mathbin{\circ}\Varid{f})\;\Varid{c}){}\<[E]%
\\
\>[B]{}={}\<[BE]%
\>[5]{}\mbox{\commentbegin  naturality of \ensuremath{\mathit{join_M}}  \commentend}{}\<[E]%
\\
\>[B]{}\hsindent{3}{}\<[3]%
\>[3]{}(\Conid{M}\;\mathit{Return}\mathbin{\circ}\mathit{join_M})\;(\Conid{M}\;\Varid{f}\;\Varid{c}){}\<[E]%
\\
\>[B]{}={}\<[BE]%
\>[5]{}\mbox{\commentbegin  definition of \ensuremath{\Varid{lift}}  \commentend}{}\<[E]%
\\
\>[B]{}\hsindent{3}{}\<[3]%
\>[3]{}\Varid{lift}\;(\mathit{join_M}\;(\Conid{M}\;\Varid{f}\;\Varid{c})){}\<[E]%
\\
\>[B]{}={}\<[BE]%
\>[5]{}\mbox{\commentbegin  definition of \ensuremath{\mathit{join_M}}  \commentend}{}\<[E]%
\\
\>[B]{}\hsindent{3}{}\<[3]%
\>[3]{}\Varid{lift}\;(\Varid{c}\;\bind_M\;\Varid{f}){}\<[E]%
\ColumnHook
\end{hscode}\resethooks

To prove the fact that \ensuremath{\mathit{drop}} is a monad morphism we first prove a lemma
\begin{hscode}\SaveRestoreHook
\column{B}{@{}>{\hspre}l<{\hspost}@{}}%
\column{E}{@{}>{\hspre}l<{\hspost}@{}}%
\>[B]{}\mathit{join_M}\mathbin{\circ}\Conid{M}\;\Varid{revert}\mathbin{\circ}\Varid{prod}\mathbin{\circ}\Conid{F}\;\Varid{f}=\mathit{join_M}\mathbin{\circ}\mathit{join_M}\mathbin{\circ}\Conid{MM}\;\Varid{revert}\mathbin{\circ}\Conid{M}\;\Varid{f}\mathbin{\circ}\Varid{revert}{}\<[E]%
\ColumnHook
\end{hscode}\resethooks
The case for \ensuremath{\mathit{Return}} follows from simple unfolding of the definitions. The case for \ensuremath{\mathit{Wrap}} is as follows.

\begin{hscode}\SaveRestoreHook
\column{B}{@{}>{\hspre}c<{\hspost}@{}}%
\column{BE}{@{}l@{}}%
\column{3}{@{}>{\hspre}l<{\hspost}@{}}%
\column{5}{@{}>{\hspre}l<{\hspost}@{}}%
\column{12}{@{}>{\hspre}l<{\hspost}@{}}%
\column{E}{@{}>{\hspre}l<{\hspost}@{}}%
\>[3]{}(\mathit{join_M}\mathbin{\circ}\Conid{M}\;\Varid{revert}\mathbin{\circ}\Varid{prod}\mathbin{\circ}\Conid{F}\;\Varid{f})\;(\mathit{Wrap}\;\Varid{m}){}\<[E]%
\\
\>[B]{}={}\<[BE]%
\>[5]{}\mbox{\commentbegin  definition of \ensuremath{\Conid{F}}  \commentend}{}\<[E]%
\\
\>[B]{}\hsindent{3}{}\<[3]%
\>[3]{}(\mathit{join_M}\mathbin{\circ}\Conid{M}\;\Varid{revert}\mathbin{\circ}\Varid{prod})\;(\mathit{Wrap}\;(\Conid{MF}\;\Varid{f})\;\Varid{m})){}\<[E]%
\\
\>[B]{}={}\<[BE]%
\>[5]{}\mbox{\commentbegin  definition of \ensuremath{\Varid{prod}}  \commentend}{}\<[E]%
\\
\>[B]{}\hsindent{3}{}\<[3]%
\>[3]{}(\mathit{join_M}\mathbin{\circ}\Conid{M}\;\Varid{revert}\mathbin{\circ}\mathit{return_M}\mathbin{\circ}\mathit{Wrap}\mathbin{\circ}\mathit{join_M}\mathbin{\circ}\Conid{M}\;\Varid{prod}\mathbin{\circ}\Conid{MF}\;\Varid{f})\;\Varid{m}{}\<[E]%
\\
\>[B]{}={}\<[BE]%
\>[5]{}\mbox{\commentbegin  naturality of \ensuremath{\mathit{return_M}}  \commentend}{}\<[E]%
\\
\>[B]{}\hsindent{3}{}\<[3]%
\>[3]{}(\mathit{join_M}\mathbin{\circ}\mathit{return_M}\mathbin{\circ}\Varid{revert}\mathbin{\circ}\mathit{Wrap}\mathbin{\circ}\mathit{join_M}\mathbin{\circ}\Conid{M}\;\Varid{prod}\mathbin{\circ}\Conid{MF}\;\Varid{f})\;\Varid{m}{}\<[E]%
\\
\>[B]{}={}\<[BE]%
\>[5]{}\mbox{\commentbegin  monad laws  \commentend}{}\<[E]%
\\
\>[B]{}\hsindent{3}{}\<[3]%
\>[3]{}(\Varid{revert}\mathbin{\circ}\mathit{Wrap}\mathbin{\circ}\mathit{join_M}\mathbin{\circ}\Conid{M}\;\Varid{prod}\mathbin{\circ}\Conid{MF}\;\Varid{f})\;\Varid{m}{}\<[E]%
\\
\>[B]{}={}\<[BE]%
\>[5]{}\mbox{\commentbegin  definition of \ensuremath{\Varid{revert}}  \commentend}{}\<[E]%
\\
\>[B]{}\hsindent{3}{}\<[3]%
\>[3]{}(\mathit{join_M}\mathbin{\circ}\Conid{M}\;\Varid{revert}\mathbin{\circ}\mathit{join_M}\mathbin{\circ}\Conid{M}\;\Varid{prod}\mathbin{\circ}\Conid{MF}\;\Varid{f})\;\Varid{m}{}\<[E]%
\\
\>[B]{}={}\<[BE]%
\>[5]{}\mbox{\commentbegin  naturality of \ensuremath{\mathit{join_M}}  \commentend}{}\<[E]%
\\
\>[B]{}\hsindent{3}{}\<[3]%
\>[3]{}(\mathit{join_M}\mathbin{\circ}\mathit{join_M}\mathbin{\circ}\Conid{MM}\;\Varid{revert}\mathbin{\circ}\Conid{M}\;\Varid{prod}\mathbin{\circ}\Conid{MF}\;\Varid{f})\;\Varid{m}{}\<[E]%
\\
\>[B]{}={}\<[BE]%
\>[5]{}\mbox{\commentbegin  monad laws  \commentend}{}\<[E]%
\\
\>[B]{}\hsindent{3}{}\<[3]%
\>[3]{}(\mathit{join_M}\mathbin{\circ}\Conid{M}\;\mathit{join_M}\mathbin{\circ}\Conid{MM}\;\Varid{revert}\mathbin{\circ}\Conid{M}\;\Varid{prod}\mathbin{\circ}\Conid{MF}\;\Varid{f}))\;\Varid{m}{}\<[E]%
\\
\>[B]{}={}\<[BE]%
\>[5]{}\mbox{\commentbegin  functor  \commentend}{}\<[E]%
\\
\>[B]{}\hsindent{3}{}\<[3]%
\>[3]{}(\mathit{join_M}\mathbin{\circ}\Conid{M}\;(\mathit{join_M}\mathbin{\circ}\Conid{M}\;\Varid{revert}\mathbin{\circ}\Varid{prod}\mathbin{\circ}\Conid{F}\;\Varid{f}))\;\Varid{m}{}\<[E]%
\\
\>[B]{}={}\<[BE]%
\>[5]{}\mbox{\commentbegin  induction  \commentend}{}\<[E]%
\\
\>[B]{}\hsindent{3}{}\<[3]%
\>[3]{}(\mathit{join_M}\mathbin{\circ}\Conid{M}\;(\mathit{join_M}\mathbin{\circ}\Conid{M}\;\mathit{join_M}\mathbin{\circ}\Conid{MM}\;\Varid{revert}\mathbin{\circ}\Conid{M}\;\Varid{f}\mathbin{\circ}\Varid{revert})\;\Varid{m}{}\<[E]%
\\
\>[B]{}={}\<[BE]%
\>[5]{}\mbox{\commentbegin  functor  \commentend}{}\<[E]%
\\
\>[B]{}\hsindent{3}{}\<[3]%
\>[3]{}(\mathit{join_M}{}\<[12]%
\>[12]{}\mathbin{\circ}\Conid{M}\;\mathit{join_M}\mathbin{\circ}\Conid{MM}\;\mathit{join_M}\mathbin{\circ}\Conid{MMM}\;\Varid{revert}\mathbin{\circ}\Conid{MM}\;\Varid{f}\mathbin{\circ}\Conid{M}\;\Varid{revert})\;\Varid{m}{}\<[E]%
\\
\>[B]{}={}\<[BE]%
\>[5]{}\mbox{\commentbegin  monad laws  \commentend}{}\<[E]%
\\
\>[B]{}\hsindent{3}{}\<[3]%
\>[3]{}(\mathit{join_M}{}\<[12]%
\>[12]{}\mathbin{\circ}\mathit{join_M}\mathbin{\circ}\Conid{MM}\;\mathit{join_M}\mathbin{\circ}\Conid{MMM}\;\Varid{revert}\mathbin{\circ}\Conid{MM}\;\Varid{f}\mathbin{\circ}\Conid{M}\;\Varid{revert})\;\Varid{m}{}\<[E]%
\\
\>[B]{}={}\<[BE]%
\>[5]{}\mbox{\commentbegin  naturality of \ensuremath{\mathit{join_M}}  \commentend}{}\<[E]%
\\
\>[B]{}\hsindent{3}{}\<[3]%
\>[3]{}(\mathit{join_M}{}\<[12]%
\>[12]{}\mathbin{\circ}\Conid{M}\;\mathit{join_M}\mathbin{\circ}\Varid{join\char95 }\;\Varid{m}\mathbin{\circ}\Conid{MMM}\;\Varid{revert}\mathbin{\circ}\Conid{MM}\;\Varid{f}\mathbin{\circ}\Conid{M}\;\Varid{revert})\;\Varid{m}{}\<[E]%
\\
\>[B]{}={}\<[BE]%
\>[5]{}\mbox{\commentbegin  naturality of \ensuremath{\mathit{join_M}}  \commentend}{}\<[E]%
\\
\>[B]{}\hsindent{3}{}\<[3]%
\>[3]{}(\mathit{join_M}{}\<[12]%
\>[12]{}\mathbin{\circ}\Conid{M}\;\mathit{join_M}\mathbin{\circ}\Conid{MM}\;\Varid{revert}\mathbin{\circ}\mathit{join_M}\mathbin{\circ}\Conid{MM}\;\Varid{f}\mathbin{\circ}\Conid{M}\;\Varid{revert})\;\Varid{m}{}\<[E]%
\\
\>[B]{}={}\<[BE]%
\>[5]{}\mbox{\commentbegin  naturality of \ensuremath{\mathit{join_M}}  \commentend}{}\<[E]%
\\
\>[B]{}\hsindent{3}{}\<[3]%
\>[3]{}(\mathit{join_M}\mathbin{\circ}\mathit{join_M}\mathbin{\circ}\Conid{MM}\;\Varid{revert}\mathbin{\circ}\Conid{M}\;\Varid{f}\mathbin{\circ}\mathit{join_M}\mathbin{\circ}\Conid{M}\;\Varid{revert})\;\Varid{m}{}\<[E]%
\\
\>[B]{}={}\<[BE]%
\>[5]{}\mbox{\commentbegin  definition of \ensuremath{\Varid{revert}}  \commentend}{}\<[E]%
\\
\>[B]{}\hsindent{3}{}\<[3]%
\>[3]{}(\mathit{join_M}\mathbin{\circ}\mathit{join_M}\mathbin{\circ}\Conid{MM}\;\Varid{revert}\mathbin{\circ}\Conid{M}\;\Varid{f}\mathbin{\circ}\Varid{revert})\;(\mathit{Wrap}\;\Varid{m}){}\<[E]%
\ColumnHook
\end{hscode}\resethooks
To prove that \ensuremath{\mathit{drop}\;(\Varid{c}\;\bind_N\;\Varid{f})=\mathit{drop}\;\Varid{c}\;\bind_M\;(\mathit{drop}\mathbin{\circ}\Varid{f})}, for \ensuremath{\Varid{c}\mathbin{::}\Varid{a}} and \ensuremath{\Varid{f}\mathbin{::}\Varid{a}\to \mathit{Nest}\;\Varid{m}\;\Varid{b}}, we unfold the definition of \ensuremath{(\bind )} and prove the equivalent equality \ensuremath{\mathit{drop}\;(\Varid{join}_{\Conid{N}}\;(\mathit{N}\;\Varid{f}\;\Varid{c}))=\mathit{join_M}\;(\Conid{M}\;(\mathit{drop}\mathbin{\circ}\Varid{f})\;\mathit{drop}\;\Varid{c})}. It follows from the commutativity of the following diagram (in Hask). For brevity, we write $\mu_N$ for the \ensuremath{\Varid{join}} of monad \ensuremath{\Conid{Nest}}. The left-most path of the diagram is equal to \ensuremath{\mathit{join_M}\mathbin{\circ}\Conid{M}\;(\mathit{drop}\mathbin{\circ}\Varid{f})\mathbin{\circ}\mathit{drop}}, while the right-most path is equal to \ensuremath{\mathit{drop}\mathbin{\circ}\mathit{join_N}\mathbin{\circ}\mathit{N}\;\Varid{f}}.

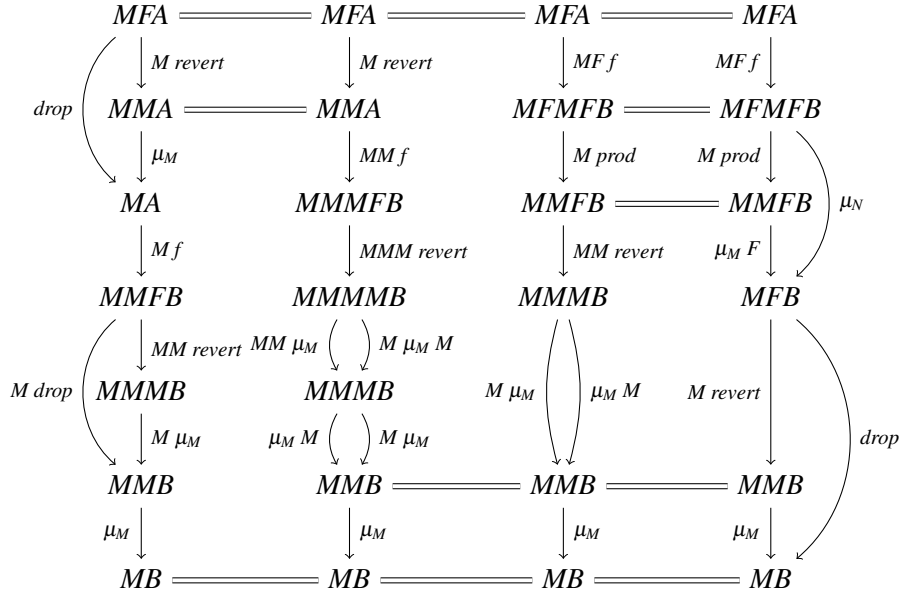
\begin{center}
\begin{tikzpicture}[description/.style={fill=white,inner sep=1pt}, rotate=90]
\matrix (m) [matrix of math nodes, row sep=1.8em,
column sep=3em, text height=1.5ex, text depth=0.25ex]
{
\ensuremath{\Conid{MFA}}&\ensuremath{\Conid{MFA}}&\ensuremath{\Conid{MFA}}&\ensuremath{\Conid{MFA}}
\\
\ensuremath{\Conid{MMA}}&\ensuremath{\Conid{MMA}}&\ensuremath{\Conid{MFMFB}}&\ensuremath{\Conid{MFMFB}}
\\
\ensuremath{\Conid{MA}}&\ensuremath{\Conid{MMMFB}}&\ensuremath{\Conid{MMFB}}&\ensuremath{\Conid{MMFB}}
\\
\ensuremath{\Conid{MMFB}}&\ensuremath{\Conid{MMMMB}}&\ensuremath{\Conid{MMMB}}&\ensuremath{\Conid{MFB}}
\\
\ensuremath{\Conid{MMMB}}&\ensuremath{\Conid{MMMB}}&
\\
\ensuremath{\Conid{MMB}}&\ensuremath{\Conid{MMB}}&\ensuremath{\Conid{MMB}}&\ensuremath{\Conid{MMB}}
\\
\ensuremath{\Conid{MB}}&\ensuremath{\Conid{MB}}&\ensuremath{\Conid{MB}}&\ensuremath{\Conid{MB}}
\\
};
\path[->,font=\scriptsize]
(m-1-1) edge node[right] {\ensuremath{\Conid{M}\;\Varid{revert}}} (m-2-1)
(m-2-1) edge node[right] {\ensuremath{\mu_M}} (m-3-1)
(m-4-1) edge node[right, yshift=-0.5mm] {\ensuremath{\Conid{MM}\;\Varid{revert}}} (m-5-1)
(m-5-1) edge node[right] {\ensuremath{M\;\mu_M}} (m-6-1)
(m-6-1) edge node[left] {\ensuremath{\mu_M}} (m-7-1)
(m-1-1) edge[bend right=50] node[left] {\ensuremath{\mathit{drop}}} (m-3-1)
(m-4-1) edge[bend right=50] node[left] {\ensuremath{M\;\mathit{drop}}} (m-6-1)
(m-3-1) edge node[right] {\ensuremath{M\;\Varid{f}}} (m-4-1)

(m-1-2) edge node[right] {\ensuremath{M\;\Varid{revert}}} (m-2-2)
(m-2-2) edge node[right] {\ensuremath{\Conid{MM}\;\Varid{f}}} (m-3-2)
(m-3-2) edge node[right] {\ensuremath{\Conid{MMM}\;\Varid{revert}}} (m-4-2)
(m-4-2) edge[bend right=30] node[left] {\ensuremath{\Conid{MM}\;\mu_M}} (m-5-2)
(m-4-2) edge[bend left=30] node[right] {\ensuremath{M\;\mu_M\;\Conid{M}}} (m-5-2)
(m-5-2) edge[bend right=30] node[left] {\ensuremath{\mu_M\;\Conid{M}}} (m-6-2)
(m-5-2) edge[bend left=30] node[right] {\ensuremath{M\;\mu_M}} (m-6-2)
(m-6-2) edge node[right] {\ensuremath{\mu_M}} (m-7-2)

(m-1-3) edge node[auto] {\ensuremath{\Conid{MF}\;\Varid{f}}} (m-2-3)
(m-2-3) edge node[auto] {\ensuremath{M\;\Varid{prod}}} (m-3-3)
(m-3-3) edge node[right] {\ensuremath{\Conid{MM}\;\Varid{revert}}} (m-4-3)
(m-4-3) edge[bend right=15] node[left] {\ensuremath{M\;\mu_M}} (m-6-3)
(m-4-3) edge[bend left=15] node[right] {\ensuremath{\mu_M\;\Conid{M}}} (m-6-3)
(m-6-3) edge node[right] {\ensuremath{\mu_M}} (m-7-3)

(m-1-4) edge node[left] {\ensuremath{\Conid{MF}\;\Varid{f}}} (m-2-4)
(m-2-4) edge node[left] {\ensuremath{M\;\Varid{prod}}} (m-3-4)
(m-3-4) edge node[left] {\ensuremath{\mu_M\;\Conid{F}}} (m-4-4)
(m-4-4) edge node[left] {\ensuremath{M\;\Varid{revert}}} (m-6-4)
(m-6-4) edge node[left] {\ensuremath{\mu_M}} (m-7-4)

(m-4-4) edge[bend left=50] node[right] {\ensuremath{\Varid{drop}}} (m-7-4)
(m-2-4) edge[bend left=50] node[right] {\ensuremath{\mu_N}} (m-4-4)
;

\draw[double distance = 1.5pt] (m-1-2) -- (m-1-3);
\draw[double distance = 1.5pt] (m-6-2) -- (m-6-3);
\draw[double distance = 1.5pt] (m-7-2) -- (m-7-3);
\draw[double distance = 1.5pt] (m-1-3) -- (m-1-4);
\draw[double distance = 1.5pt] (m-2-3) -- (m-2-4);
\draw[double distance = 1.5pt] (m-3-3) -- (m-3-4);
\draw[double distance = 1.5pt] (m-6-3) -- (m-6-4);
\draw[double distance = 1.5pt] (m-7-3) -- (m-7-4);
\draw[double distance = 1.5pt] (m-1-1) -- (m-1-2);
\draw[double distance = 1.5pt] (m-2-1) -- (m-2-2);
\draw[double distance = 1.5pt] (m-7-1) -- (m-7-2);

\end{tikzpicture}
\end{center}

\noindent
The first column of the diagram depicts unfolding of the definitions of \ensuremath{\mathit{drop}}. The second column is equal to the first one due to the fact that \ensuremath{\mathit{join_M}} is a natural transformation: the morphism \ensuremath{\mathit{join_M}\mathbin{::}\Conid{MMA}\to \Conid{MA}} form the first column ``travelled'' down the path to settle as \ensuremath{\mathit{join_M}\mathbin{::}\Conid{MMMB}\to \Conid{MMB}}. Additionally, due to the monad law for \ensuremath{\mathit{join_M}}, we can exchange \ensuremath{\Conid{MM}\;\mathit{join_M}} with \ensuremath{\Conid{M}\;\mathit{join_M}}. This allows us to use the lemma, the \ensuremath{\Conid{M}}-image of which commutes columns 2 and 3. Again, we can use the monad law to change \ensuremath{\Conid{M}\;\mathit{join_M}} into \ensuremath{\mathit{join_M}}. We use the naturality of \ensuremath{\mathit{join_M}} to switch its position with the mapping of \ensuremath{\Varid{revert}}, which justify the fourth column.

\end{document}